\documentclass[12pt,aps,prl,reprint,twocolumn,superscriptaddress]{revtex4-2}

\usepackage{amsmath,amsthm,amssymb,amsfonts}
\usepackage{mathtools}
\usepackage{bm}
\usepackage{graphicx,color}
\usepackage[colorlinks=true,allcolors=blue]{hyperref}
\usepackage{xfrac}

\usepackage[normalem]{ulem}
\usepackage{sidecap}

\definecolor{jfcolor}{rgb}{0.1, 0.0, 0.9}

\definecolor{szlcolor}{rgb}{0.75, 0, 0.75}

\definecolor{mmcolor}{rgb}{0.6, 0.3, 0.0}

\begin{document}

\title{Tissue fluidization by cell-shape-controlled active stresses}
\author{Shao-Zhen Lin}
\affiliation{Aix Marseille Universit\'{e}, Universit\'{e} de Toulon, CNRS, Centre de Physique Th\'{e}orique, Turing Center for Living Systems, Marseille, France}
\author{Matthias Merkel}
\affiliation{Aix Marseille Universit\'{e}, Universit\'{e} de Toulon, CNRS, Centre de Physique Th\'{e}orique, Turing Center for Living Systems, Marseille, France}
\author{Jean-Fran\c{c}ois Rupprecht}
\affiliation{Aix Marseille Universit\'{e}, Universit\'{e} de Toulon, CNRS, Centre de Physique Th\'{e}orique, Turing Center for Living Systems, Marseille, France}

\begin{abstract}
Biological cells can actively tune their intracellular architecture according to their overall shape. Here we explore the rheological implication of such coupling in a minimal model of a dense cellular material where each cell exerts an active mechanical stress along its axis of elongation. 
Increasing the active stress amplitude leads to several transitions. An initially hexagonal crystal motif is first destabilized into a solid with anisotropic cells. Increasing activity further, we find a re-entrant transition to a regime with finite hexatic order and finite shear modulus, in which cells arrange according to a rhombile pattern with periodically arranged rosette structures. The shear modulus vanishes again at a third threshold beyond which spontaneous tissue flows arise. In this last regime, we observe the emergence of cell shape patterns called topological defects, with flow and stress fields around defects agreeing with those observed in epithelial tissue experiments. We further provide a testable prediction of cell-cell rearrangement hotspots near topological defects. 
Overall, our work connects seemingly distinct features -- e.g.\ rosettes and topological defects -- observed across various types of epithelial tissues.
\end{abstract} 
\date{\today}

\maketitle

Connecting the single-cell behavior to large-scale mechanical properties of biological tissues is key to understand development, regeneration, and disease \cite{Ladoux2017}. Growing experimental evidence supports the idea that biological cells actively tune their intracellular architecture according to their overall shape. For instance, cortical actomyosin  \cite{Blanchoinreview,Gupta2019,Gorelova2021} and microtubules (whether in \textit{Drosophila} \cite{dye2021self,Singh2018} or in plants \cite{Mirabet2018,Gorelova2021}) tend to align along the direction of cell shape elongation. Such oriented fibers are known to generate anisotropic stress \cite{Prost2015}. 

Here, we explore the consequences of such cell-shape feedback on the tissue-scale behavior.
We study a minimal model where a bulk cellular active stress $\bm{\sigma}^{(\rm act)}$ is created by filaments, which in turn align with cell shape, represented by a tensor $\bm{Q}$. To lowest order, 
\begin{equation} 
\bm{\sigma}^{(\rm act)} = -\beta \bm{Q}, \label{eq_Shape_ActiveStress_Coupling}
\end{equation}
which we incorporate within a computational model for  dense epithelilal tissue. For $\beta > 0$ (resp.\ $\beta<0$), cells actively push (resp.\ pull) on their neighbours along their direction of elongation \cite{Duclos_NP_2018,Saw_Nature_2017}. 

Relations like Eq.~(\ref{eq_Shape_ActiveStress_Coupling}) have been considered in previous models. For instance, in active hydrodynamic theories, where both $\bm{\sigma}^{(\rm act)}$ and $\bm{Q}$ are defined by averages over several cells, Eq.~(\ref{eq_Shape_ActiveStress_Coupling}) gives rise to a classical flow instability \cite{Simha2002,Voituriez2005,Marchetti2013}. In the absence of confining boundaries, this instability occurs at arbitrarily small activities $\beta >\beta_c\equiv 0$.
An active stress as in Eq.~\eqref{eq_Shape_ActiveStress_Coupling} is also included in cell-based phase field simulations \cite{Mueller2019}, where $\bm{Q}$ is defined at the single-cell level. In these simulations the transition to spontaneous flows occurs for a critical value of $\beta_c$, which (i) is finite, $\beta_c>0$, and appears to be independent of the system size, contrasting with active hydrodynamic theory results, and (ii) scales with a cell surface tension. 
So far it is unclear why the critical activity $\beta_c$ is different in these two models, and why it scales with the surface tension in the phase field model Ref.~\cite{Mueller2019}.

Here we address these questions by combining analytical arguments and vertex model simulations. Vertex models describe epithelial tissues as networks of polygons \cite{Alt2017}. Forces on the polygon vertices are defined by a mechanical energy: $E = 1/2\sum_{n}K_A(A_n-A_0)^2 + K_P(P_n-P_0)^2$
where the sum is over all cells $n$ of the tissue, $A_n$ and $P_n$ are cell area and perimeter, respectively. The parameters $A_0$ and $P_0$ are preferred cell area and perimeter with the associated rigidities $K_A$ and $K_P$, respectively.  A transition occurs at $P_0=P_0^\ast$, with the tissue behaving as a yield stress solid for $P_0<P_0^\ast$ and as a fluid for $P_0\geq P_0^\ast$ \cite{Farhadifar2007,Bi2015}. The numerical value of $P_0^\ast$ is in the range $3.72\dots3.94$, which a value that depends on the disorder in the cellular packing \cite{Farhadifar2007,Bi2015,Sussman2018,Merkel2019,Wang2020} (SM \cite{SM}, Sec. I). Cell-based active polar forces at the individual cell level were also shown to drive a solid-to-fluid transition through an intermediate hexatic regime \cite{Bi2016,Pasupalak2020}.

\begin{SCfigure*}
\centering
\includegraphics[width=14cm]{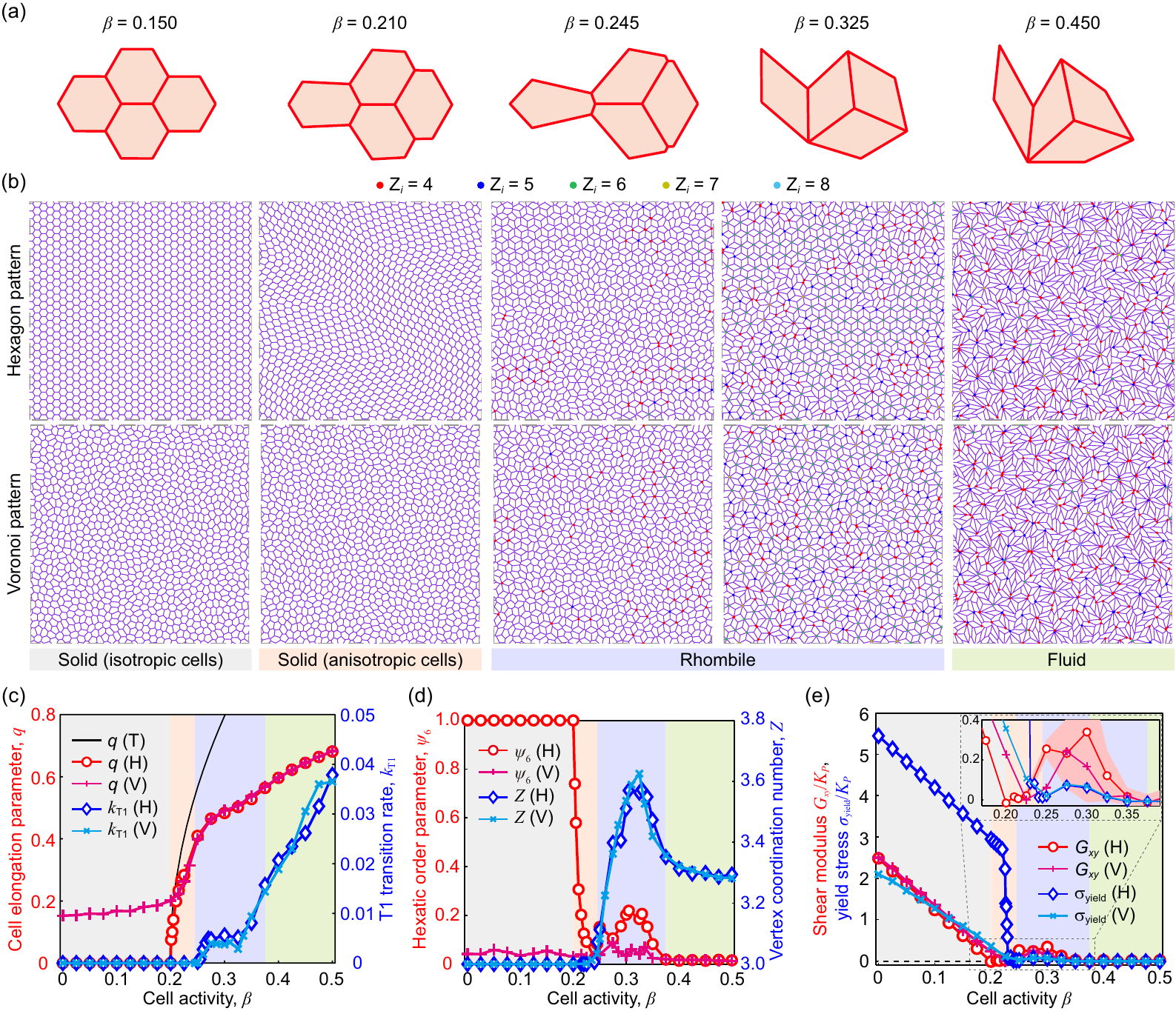}
\caption{\label{fig_1} 
(a, b) Morphology of (a) a four-cell system and (b) a cell sheet at different activities $\beta$, with an initially hexagon (top) or Voronoi pattern (bottom). 
In (b), we mark vertices with coordination number $Z_i > 3$. 
(c) Cell elongation parameter $q$, Eq.~(\ref{eq:Q}), and T1 transition rate $k_{\rm T1}$ versus cell activity $\beta$, for an initially hexagonal (H) or Voronoi (V) pattern. The solid black line refers to an analytical approximation of $q$.
(d) Hexatic order parameter $\psi_6$ and average vertex coordination number $Z$ versus cell activity $\beta$. 
(e) Long-time shear modulus $G_{xy}$ and yield stress $\sigma_{\rm yield}$ versus cell activity $\beta$. 
Inset: red shaded area indicates the mean $\pm$ standard deviation of the shear modulus for the initially hexagonal pattern, estimated from $n=5$ simulations. 
Default parameter values with $P_0 = 1.0$. }
\end{SCfigure*}

We introduce Eq. (\ref{eq_Shape_ActiveStress_Coupling}) in the vertex model framework, and we find that three  transitions occur with increasing $\beta$. For $\beta < \beta_1$ the tissue is solid with isotropic cells. At $\beta = \beta_1$, cells lose their isotropic shape. At a second transition $\beta=\beta_2$, a regime with rhombile cell shapes, many-fold vertices, long-range crystalline order and finite shear modulus emerges. Finally, at 
$\beta = \beta_3$, the tissue turns into an active fluid and starts to display persistent small-scale chaotic flow.
We discuss the relation of these transitions to the $P_0$-dependent vertex model transition between yield stress solid and fluid. 
This also allows us to demonstrate the appearance of a finite activity threshold $\beta_c$ in cell-based models such as Ref.~\cite{Mueller2019};  the latter behave like yield stress solids, requiring finite $\beta_c$ for active flows to appear. Meanwhile, active hydrodynamic theories as in \cite{Simha2002,Voituriez2005} describe tissues as fluids, such that active flows can appear without threshold. We then discuss potential applications to the modelling of biological tissues.

\textbf{Method} We implement Eq.~\eqref{eq_Shape_ActiveStress_Coupling} through the following friction-based dynamics for the vertex positions $\bm{r}_i$
\begin{align}
\gamma \frac{\mathrm{d} \bm{r}_i}{\mathrm{d} t} = \bm{F}_i^{(\rm svm)} + \bm{F}_i^{(\rm act)}. \label{eq:dynamics}
\end{align} 
Here, $\gamma$ is a friction coefficient,  $\bm{F}_i^{(\rm svm)} = -\partial E / \partial \bm{r}_i$ are the standard gradient-based vertex model forces, and $\bm{F}_i^{(\rm act)}$ are the active forces induced by the active anisotropic bulk stresses $\bm{\sigma}^{(\rm act)}$ defined for each cell according to Eq.~\eqref{eq_Shape_ActiveStress_Coupling}. There are different ways to translate the cellular bulk stresses $\bm{\sigma}^{(\rm act)}$ into the vertex forces $\bm{F}_i^{(\rm act)}$ \cite{Tlili_PNAS_2019,Comelles_eLife_2021,Lin2022}. Here we use the approach proposed by Tlili et al.\ \cite{Tlili_PNAS_2019,Lin2022}, which relies on Cauchy’s stress definition.
For the cell shape anisotropy tensor $\bm{Q}$ in Eq.~\eqref{eq_Shape_ActiveStress_Coupling}, we use the symmetric, traceless tensor 
\begin{equation} \label{eq:Q}
    \bm{Q} = (\sum_k{{{l}_{k}}{{\bm{t}}_{k}}\otimes {{\bm{t}}_{k}}})/P-\bm{I}/2,
\end{equation}
where the sum is over all sides $k$ of the cell, while $l_k$ and $\bm{t}^k$ denote length and unit tangent vector of side $k$, respectively.  The eigenvalues and principal directions of $\bm{Q}$ provide metrics for the cell shape and cell orientation. We measure the cell shape anisotropy by $q = \sqrt{2{\rm tr}(\bm{Q}^2)} \in[0,1)$, with $q=0$ for round cells and $q\rightarrow1$ for increasingly elongated ones. 

We initialize the system with cells arranged according to either (i) a regular hexagonal pattern with small random deviations in the vertex positions, or (ii) random Voronoi tessellations (SM \cite{SM}, Sec. I). We use periodic boundary conditions with fixed system size. We set $K_A=1$, $A_0=1$, $K_P=0.02$, $\gamma=1$, $N = 10^3$ cells and $P_0=1$ if not otherwise stated. 

\begin{figure}[t!]
\centering
\includegraphics[width=8.6cm]{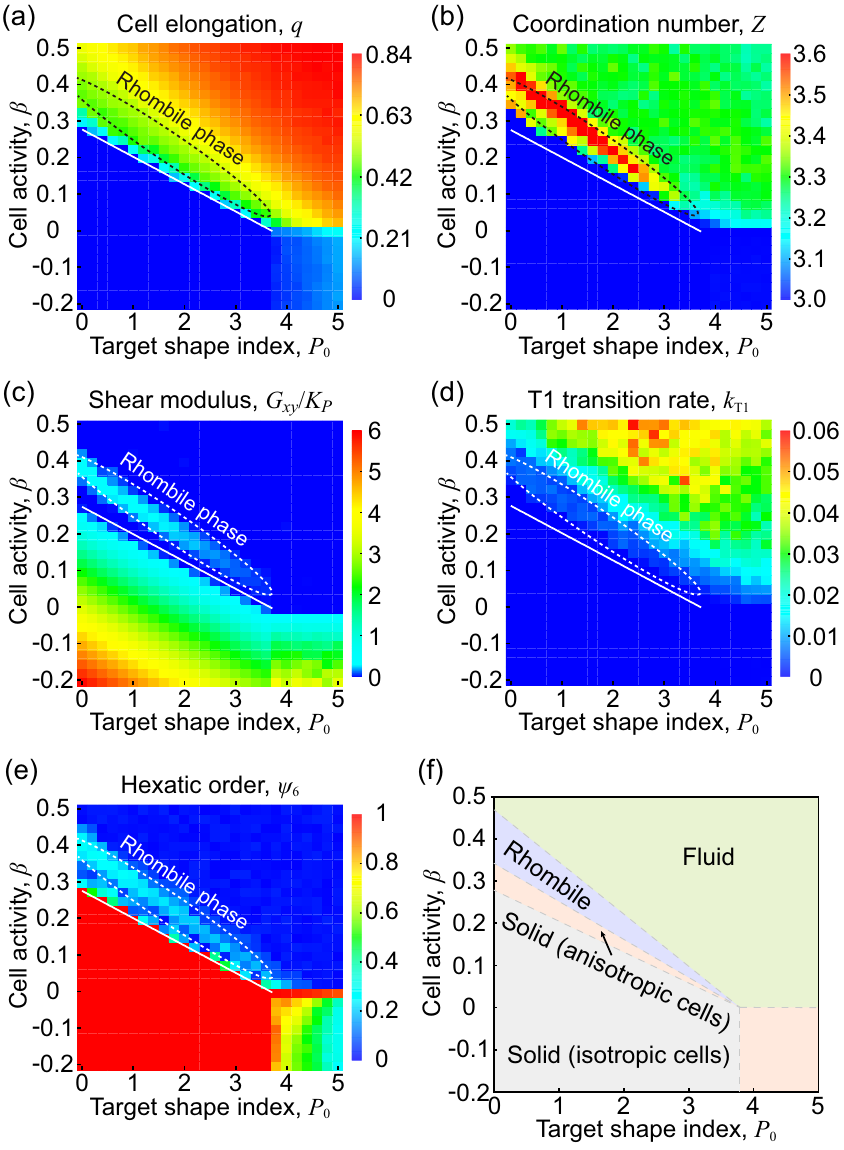}
\caption{\label{fig_2} Diagrams depending on the target shape index $P_0$ (inter-cellular tension) and cell bulk activity $\beta$, showing (a) average cell elongation parameter $q$, (b) average vertex coordination number $Z$, (c) shear modulus $G_{xy}$, (d) T1 topological transition rate $k_{\rm T1}$, and (e) hexatic order parameter $\psi_6$. Here, the white solid curves refers to the theoretical prediction $\beta_1 (P_0)$, Eq.~(\ref{eq:beta_1}), for the transition between the solid regime with isotropic cells and the solid regime with anisotropic cells. (f) Overall phase diagram. }
\end{figure}

\textbf{Results} Increasing $\beta$, we observe several rheological and structural transitions (Fig.~\ref{fig_1}a,b; Movies S1-S3). In Fig.~\ref{fig_2} we show the dependence of these transitions on both $\beta$ and $P_0$; however, in the following, we focus on the case $P_0=1$ (Fig.~\ref{fig_1}). 
For small $\beta<\beta_1\approx 0.20$ the vertex model tissue is solid (Fig.~\ref{fig_1}e) with isotropic cell shapes (Fig.~\ref{fig_1}c), where the average cell elongation is $q=0$ (resp.\ $q\approx 0.16$) when using a hexagonal (resp.\ Voronoi) initial state. 
When $\beta$ increases beyond $\beta_1$, cell shapes become anisotropic, as indicated by an increase in $q$ (Fig.~\ref{fig_1}c). This is accompanied by a decrease in the hexatic order parameter $\psi_6$ (Fig.~\ref{fig_1}d), defined as $\psi_6 = \left| \sum {{{\Psi }_{j}}} / N \right|$, where ${{\Psi }_{j}} = \sum_{k\in \text{neighbors}}{\exp \left( \text{i}6{{\theta }_{jk}} \right)} / N_j$ and $\theta_{jk} = \arg \left( \bm{r}_k - \bm{r}_j \right)$ \cite{Li2018,Paoluzzi2021}. While in this regime the shear modulus vanishes for the hexagonal initial state, the tissue remains solid, as verified through the examination of the yield stress (Fig.~\ref{fig_1}e, SM \cite{SM}, Sec. I). The transition point $\beta_1$ decreases with increasing $P_0$ up until the critical point $P_0^\ast$ (Fig.~\ref{fig_2}).

The cell shape transition at $\beta = \beta_1$ occurs because the $\beta$ term in Eq.~\eqref{eq_Shape_ActiveStress_Coupling} effectively corresponds to a negative shear modulus. As a consequence, when $\beta>\beta_1$, the total cellular shear modulus decreases to zero,  destabilizing the isotropic cell shape.  To see this, we start from the Batchelor stress of a vertex model cell with perimeter $P$ and area $A$ \cite{Batchelor_JFM_1970,Lau_PRE_2009,Nestor-Bergmann_MMB_2018,Lin2022}, whose anisotropic part $\bm{\tilde\sigma}$ is (SM \cite{SM}, Sec. I):
\begin{equation}
  \bm{\tilde\sigma} = \left[\frac{K_PP(P-P_0)}{A} - \beta\right]\bm{Q}.\label{eq:tilde sigma mean field}
\end{equation}
To obtain the global tissue shear modulus $G_\mathrm{aff}$ in an analytical mean-field picture, we apply an affine pure shear strain $\epsilon$ to an isotropic cell, which creates the cell shape anisotropy $q = 3\epsilon/2$ to lowest order in $\epsilon$ (SM \cite{SM}, Sec. II).  Comparing Eq.~\eqref{eq:tilde sigma mean field} to $\tilde\sigma=2G_\mathrm{aff}\epsilon$, we obtain:
\begin{equation}
  G_\mathrm{aff} = \frac{3}{8}\Big[K_PP(P-P_0)-\beta\Big].\label{eq:G_aff}
\end{equation}
Here, we have used $A\approx A_0=1$, which corresponds to the limit of incompressible cells.
Testing Eq.~\eqref{eq:G_aff} numerically, where we also include all non-affinities, we find the same result, except for a prefactor:  $G_\mathrm{non-aff}\approx2G_\mathrm{aff}/3$.
To determine the perimeter $P$ appearing in Eq.~\eqref{eq:G_aff}, we use that in the isotropic solid regime $P=P_0^\ast\approx 3.722$ for a hexagonal tissue \cite{Staple2010,Bi2015}.
Isotropic cell shape thus becomes unstable for $\beta>\beta_1(P_0)$ with (SM \cite{SM}, Sec. II):
\begin{equation}
\beta_1(P_0) = K_P P_0^\ast(P_0^\ast-P_0).\label{eq:beta_1}
\end{equation}
This equation exactly predicts the stability of the regular hexagonal crystal (white lines in Fig.~\ref{fig_2}). 

This mean-field picture also explains how in the regime $P_0<P^{\star}_0$ cells elongate for $\beta>\beta_1(P_0)$. For an affinely sheared isotropic cell, the perimeter increases quadratically with its shape anisotropy as $P=P_0^\ast(1+q^2/3)$ (SM \cite{SM}, Sec. II) \cite{Wang2020}. Inserting this in Eq.~\eqref{eq:tilde sigma mean field} and combining it with $\tilde\sigma=(\partial E/\partial\epsilon)/2A$, we obtain an effective potential of the cell depending on its shape anisotropy $E_{\rm eff}(q)$, which reads to fourth order in $q$ (SM \cite{SM}, Sec. II): 
\begin{equation} \label{eq:effective_energy}
E_{\rm eff}(q) = \frac{1}{3}\Big[\beta_1(P_0)-\beta\Big]q^2 + \frac{1}{18}K_P P_0^\ast\big(2P_0^\ast-P_0\big)q^4 . 
\end{equation}
The energy minimum for $\beta<\beta_1$ is at $q_\mathrm{min}=0$, while for $\beta>\beta_1$ the minimum is at $q_{\min}(\beta) = \sqrt{3(\beta-\beta_1)/K_PP_0^\ast(2P_0^\ast-P_0)}$, which corresponds to a typical pitchfork bifurcation.
Indeed, this predicts well the observed cell elongation in the regime close to $\beta_1$ for $P<P^{\star}_0$ (Fig.~\ref{fig_1}c, SM \cite{SM}, Sec. II). 

The behavior of cell-shape elongation $q$ is different in the regime $P_0>P^{\star}_0$, where we observe a \textit{discontinuous} increase in $q$ and coordination number $Z$ as soon as $\beta$ is increased above zero (Fig.~\ref{fig_2}a,b, SM \cite{SM}, Sec.~VI). 
The discontinuity in $q$ can be understood by noting that in the regime $P_0>P^{\star}_0$ the standard vertex model is floppy with vanishing energy $E(q)=0$ until some critical $q$ value $q_\mathrm{crit}\sim\sqrt{P_0-P_0^\ast}$, beyond which cells and the vertex model tissue start to attain a finite shear modulus \cite{Wang2020}.
As a consequence, as soon as $\beta$ is set to a positive value, the energy below $q_\mathrm{crit}$ becomes $E(q)\sim-\beta q^2$ and so any state $q<q_\mathrm{crit}$ becomes unstable.
Thus, for small positive $\beta$, cell elongation will make a jump from zero to a value close to $q_\mathrm{crit}$.

For $P_0<P_0^\ast$, the prediction for $q$ in Eq.~\eqref{eq:effective_energy} breaks down beyond some $\beta > \beta_2$, where $\beta_2\approx 0.24$ for $P_0=1$ (Fig.~\ref{fig_1}c). This corresponds to a regime marked by an increasing hexatic order, which, for tissues initiated in the near hexagonal pattern, appears to be even long-ranged  (SM \cite{SM}, Sec. I). (Fig.~\ref{fig_1}b-d). We call this the \emph{rhombile} regime because domains appear where cells attain a rhombic shape, which arrange into a periodic arrangement of six-fold vertices (Fig.~\ref{fig_1}b).  Correspondingly, the rhombile regime is marked by the increase in the average vertex coordination number $Z$ (Fig.~\ref{fig_1}d). 
We observe such behavior within an intermediate range of activities $\beta$ for all $P_0 < P^\ast_0$, as revealed through an increase in $Z$ and $\psi_6$ (Fig.~\ref{fig_2}b,e).

The emergence of the rhombile pattern and manyfold vertices can be understood from single- and four-cell systems. Indeed, at a critical value of $\beta^{(1)}_2 \approx 0.27$ for the single-cell system ($\beta^{(4)}_2 \approx 0.25$ for the four-cell system), the two shortest cell edges shrink to length zero, resulting in the observed rhombic cell shapes (Fig.~\ref{fig_1}a; Movie S4-S5; SM \cite{SM} Sec. II). 

We find that the shear modulus in the rhombile regime is finite and peaks when the rhombile domain extension is maximal at around $\beta^{*} \approx 0.3$, as reflected by the local maximum in the hexatic order $\psi_6$ and coordination number $Z$ (Fig.~\ref{fig_1}d). 
We understand this by considering a single cell: at $\beta^{*} = 0.3$, the cell reaches a regular diamond shape with two $\pi/3$ and two $2\pi/3$ angles. This shape (also called \textit{calisson} \cite{Alsina2015}) is the building block of the crystal rhombile pattern (SM \cite{SM}, Fig.~S10b). When $\beta < \beta^{*} = 0.3$ (resp.\ $\beta > \beta^{*} = 0.3$), acute angles above (resp.\ below) $\pi/3$ form in the single cell, leading to frustration in the crystal rhombile pattern, which can destabilize the rhombile domains. 

Despite a finite shear modulus, the rhombile regime exhibits a finite steady-state T1 transition rate (Fig.~\ref{fig_1}d; SM \cite{SM} Sec. III; Movie S2). This may appear paradoxical at first sight, since T1 transitions are expected to relax the applied stress \cite{Marmottant2009}, leading to a long-time fluid material response. While such argument is true in a passive material at equilibrium, it may not hold for an active system at steady state, as observed here. Further, we notice that these T1 transitions are generated at the interfaces, rather than in the bulk, of the rhombile crystal domains (Movie~S2). 

The shear modulus eventually vanishes at $\beta_3 \approx 0.37$. We find that this value matches the one at which the shear modulus of the rhombile crystal vanishes (SM \cite{SM}, Sec. III). This suggests that the finite shear modulus of the rhombile regime is created by the rhombile crystal domains.

\begin{figure}[t!]
\centering
\includegraphics[width=8cm]{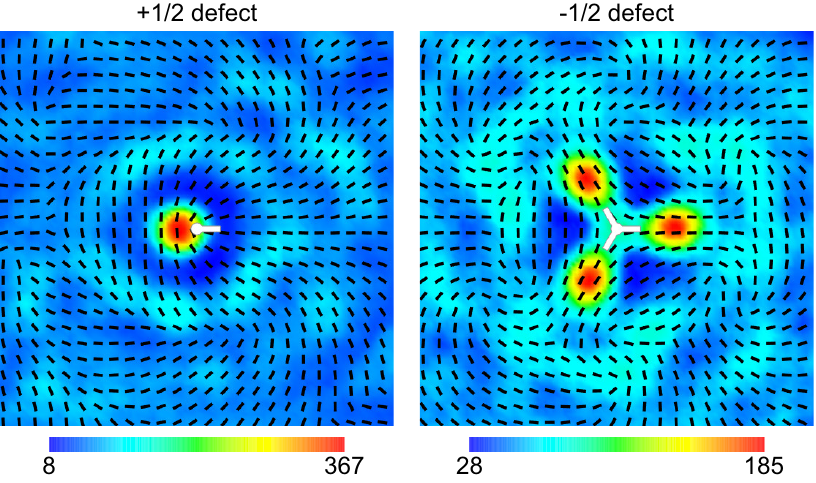}
\caption{\label{fig_3} Average T1 transition rate per unit area (color bar) within a box of side length 10 centered around the core of $+1/2$ (left) and $-1/2$ (right) topological defects. We also applied rotations to align the defect orientations. Black lines indicate the local average of cell orientation. Default parameter values with $P_0=1$ and $\beta = 0.4$; average over $n = 9844$ half-integer defects.}
\end{figure}

For $\beta > \beta_3$, the system flows continuously, where the steady-state T1 transition rate increases with activity (Fig.~\ref{fig_1}e, \ref{fig_2}d and Movie S2). Such flow already appears in a four-cell system for an intermediate activity range (Movie~S6). 
This flowing regime exhibits features of an active nematic material. For instance, coarse-graining of the cell orientation field reveals the presence of $\pm 1/2$ topological defects in the fluid regime (SM \cite{SM}, Sec. V) \cite{DeCamp_NM_2015,Vromans_SoftMatter_2016}. The stress and velocity patterns around $\pm 1/2$ topological defects are consistent with those predicted in an incompressible material with extensile nematic activity \cite{Marchetti2013,Giomi2015,Mueller2019} (SM \cite{SM}, Sec. V). Our vertex model simulations also allow us to make new predictions at the cell scale. For instance, we find that close to the $\pm 1/2$ defect cores there are hotspots of T1 transitions (Fig.~\ref{fig_3}, SM \cite{SM}, Sec. V). 

We are now in the position to explain the apparent contradiction between the previous cell-based model \cite{Mueller2019}, showing the onset of active flows only beyond a finite critical activity $\beta_c > 0$, and active continuum models \cite{Voituriez2005,Simha2002}, where active flows appear at any finite activity, i.e.\ $\beta_c=0$. In our vertex model, such a difference arises as a function of $P_0$. For $P_0<P^{\star}_0$, where the vertex model behaves as yield stress solid, we showed that $\beta$ acts as a negative shear modulus, destabilizing the tissue beyond a critical value that increases with the distance to $P^{\star}_0$. In contrast, we observe that for the fluid vertex model regime, $P_0>P^{\star}_0$, any positive $\beta$ destabilizes the tissue, inducing a sharp increase in flow (Fig.~\ref{fig_2}d, SM \cite{SM}, Sec. VI). Comparing both cases, we conclude that while any positive activity leads to flows in a fluid material, a yield stress creates a finite activity threshold to flow. This is likely the case in Ref.~\cite{Mueller2019}, because the employed phase field model essentially describes a foam that likely displays a yield stress.

Finally, we tested how much our results depend on the definition of the $\bm{Q}$ tensor used in Eq.~\eqref{eq_Shape_ActiveStress_Coupling} (SM \cite{SM}, Sec. IV). We find that the transitions in cell shape and rigidity are generic, but not the appearance of the rhombile crystal domains. We further observe that for two different $\bm{Q}$ definitions chevron patterns of local smectic order for the cell orientation emerge within the solid regime with anisotropic cells, $\beta_1 < \beta < \beta_2$.

\textbf{Conclusion} Here, we studied how a feedback of cell shape on the cellular active stress generation affects collective tissue dynamics.  We show that such a feedback can fluidify the vertex model tissue through a series of intermediate steps displaying hexatic order, finite shear modulus and spontaneous T1 transitions. 

\textbf{Perspectives} We anticipate several applications to the understanding of cellular materials. First, in most instances, active cellular materials have been studied in the vertex model through the introduction of traction forces against a resting substrate \cite{Giavazzi2018,Lin_BJ_2018}. 
In contrast, we introduce activity through momentum-conserving forces $\bm{F}_i^{(\rm act)}$ \cite{Lin2022}. This kind of activity could be useful to describe tissue flows in conditions of low environmental friction e.g.\ free-standing tissues, floating mouse embryos or intestinal organoids \cite{Khalilgharibi2019}. Second, our work also connects through a single active stress parameter seemingly distinct features encountered in epithelial tissues, namely rosettes \cite{Sun2017,Wang2020} and topological defects \cite{Saw_Nature_2017}. We look forward to test our prediction of T1 transition hotspots near topological defects from experimental data of epithelial tissue.

\textbf{Acknowledgements}
We thank the Centre Interdisciplinaire de Nanoscience de Marseille (CINaM) and the Laboratoire Adhésion Inflammation (LAI) for providing office space. We are funded by the Investissements d’Avenir French Government program managed by the French National Research Agency (ANR-16-CONV-0001, ANR-20-CE30-0023 COVFEFE) and by the Excellence Initiative of Aix-Marseille University - A*MIDEX.
\vskip-0.5cm


%

\end{document}


\title{Supplemental Material \\ Tissue fluidization by cell-shape-controlled active stresses}

\author{Shao-Zhen Lin}
\affiliation{Aix Marseille Universit\'{e}, Universit\'{e} de Toulon, CNRS, Centre de Physique Th\'{e}orique, Turing Center for Living Systems, Marseille, France}
\author{Matthias Merkel}
\affiliation{Aix Marseille Universit\'{e}, Universit\'{e} de Toulon, CNRS, Centre de Physique Th\'{e}orique, Turing Center for Living Systems, Marseille, France}
\author{Jean-Fran\c{c}ois Rupprecht}
\affiliation{Aix Marseille Universit\'{e}, Universit\'{e} de Toulon, CNRS, Centre de Physique Th\'{e}orique, Turing Center for Living Systems, Marseille, France}

\renewcommand\thefigure{S\arabic{figure}}   
\renewcommand{\figurename}{FIG.}   
\setcounter{figure}{0}


\date{\today}
\maketitle

\tableofcontents

\newpage
\clearpage

\section{Simulation implementation} 

\subsection{Simulation setup and parameter values} \label{sec:simulationsetup}

The cell dynamics is dictated by the motions of vertices, which obey the force balance equation (Eq. (\blue{2}) in the main text). In addition, we implemented T1 transitions once the length of a cell edge decreases below a critical value $\Delta_{\rm T1}$  \cite{Fletcher_BJ_2014,Bi2015,Lin_BJ_2018}. 

\begin{figure*}[h!]
 \centering
 \ifthenelse{ \cverbose > 0}{}{\includegraphics[width=15cm]{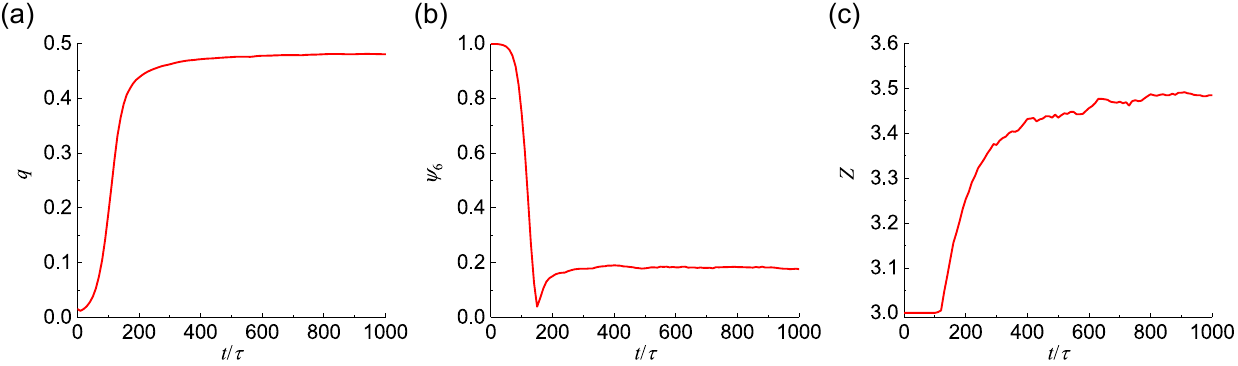} 
 }
 \caption{\label{fig_SteadyCheck} 
An example to check the dynamic steady state. Evolutions of (a) the cell elongation parameter $q$, (b) the hexatic order parameter $\psi_6$, and (c) the average vertex coordination number $Z$. 
Parameters: $K_P = 0.02$, $P_0 = 1.0$ and $\beta = 0.3$. 
}
\end{figure*}

Simulations were started from a regular hexagonal pattern (or random Voronoi pattern if stated) with small Gaussian
perturbations (magnitude = 0.01) applied to the vertices’ positions at $t = 0$, and run for enough steps (typically $N_{\rm simulation} = 100,000$) to arrive dynamic steady state, which can be checked by examining the evolutions of the cell elongation parameter $q$, the hexatic order parameter $\psi_6$, the average vertex coordination number $Z$, see Fig. \ref{fig_SteadyCheck} for example. We systematically consider system prepared with zero residual stress (see Sec. \ref{sec:initialization}). 

In our simulations, the governing equations and corresponding parameters are rescaled by the length scale $\ell =\sqrt{{{A}_{0}}}$, the time scale $\tau ={\gamma }/{\left( {{K}_A}{{A}_{0}} \right)}\;$, and the stress scale $\sigma ={{K}_A}{{A}_{0}}$.  The following non-dimensional parameter values are adopted: ${K}_P=0.02$, ${{\Delta }_{\text{T}1}}=0.01$, and $\Delta t=0.01$. If not stated otherwise, we consider a system of $N_{\rm cell} = 1,000$ and an initially hexagonal cell pattern.  In Sec. \ref{sec:Rhombile_DifferentSystemSize}, we verified that our results were consistent for a wide range of system sizes.

\subsection{Generation of a global stress-free cell sheet} \label{sec:initialization}

The stress-free configuration is defined as a configuration where the global tissue stress is zero. We here give the details of how to generate the stress-free configuration for a cell sheet without active stress. 

Once we set up a hexagonal cell pattern or random Voronoi cell pattern, we compute the global stress in the cell sheet. It is possible to compute cell-based stresses from the forces applied on the vertices using the Batchelor formula \cite{Batchelor_JFM_1970,Lau_PRE_2009,Nestor-Bergmann_MMB_2018}. In the absence of active stress, the cell stress can be expressed as: 
\begin{equation}
\bm{\sigma }_J =\left[ {{K}_{A}}\left( A_J-{{A}_{0}} \right)+\frac{1}{2}\frac{{{K}_{P}}P_J\left( P_J-{{P}_{0}} \right)}{A_J} \right]\bm{I}+\frac{{{K}_{P}}P_J\left( P_J-{{P}_{0}} \right)}{A_J}\bm{Q}_J
 , \label{eq_StressFormula_Passive}
\end{equation}
where the first term is the isotropic stress and the second term is the anisotropic stress. The global stress in a cell sheet can be calculated by, ${{\bm{\sigma }}^{\left( \text{tissue} \right)}} \left[ \{ \bm{r}_i\} \right] = {\sum_{J}{{{A}_{J}}{{\bm{\sigma }}_{J}}}} / {\sum_{J}{{{A}_{J}}}}$. 

The initial tissue state correspond to a global stress $\bm{\sigma}^{\rm (tissue)}$ that is non-zero. Here we propose a two-step protocol to achieve a near zero-stress state. 
\begin{enumerate}
    \item As a first step, we apply a uniform affine transformation to the cell sheet, $\{ \bm{r}_i \} \rightarrow \{ \lambda\bm{r}_i \}$, where $\lambda > 0$ here is a scaling factor to be determined in the following. In the new configuration, $\{ \lambda\bm{r}_i \}$, the global stress in the tissue can be expressed as 
\begin{equation}
{{\bm{\sigma }}^{\left( \text{tissue} \right)}}\left[ \left \{ \lambda {{\bm{r}}_{i}} \right \} \right] = \frac{\sum\limits_{J}{\left\{ \left[ \lambda {{K}_{A}}{{A}_{J}}\left( {{\lambda }^{2}}{{A}_{J}}-{{A}_{0}} \right)+\dfrac{1}{2}{{K}_{P}}{{P}_{J}}\left( \lambda {{P}_{J}}-{{P}_{0}} \right) \right]\bm{I}+{{K}_{P}}{{P}_{J}}\left( \lambda {{P}_{J}}-{{P}_{0}} \right){{\bm{Q}}_{J}} \right\}}}{\lambda \sum\limits_{J}{{{A}_{J}}}} .
\end{equation}
To get the stress-free state, we first reach a traceless stress state through a rescaling factor $\lambda$ that solves the condition $\mathrm{tr}\{\bm{\sigma}^{(\rm tissue)}\big[\lambda\bm{r}_i\big]\} = 0$, which reads 
\begin{equation}
\left( 4\sum\limits_{J}{\tilde{A}_{J}^{2}} \right){{\lambda }^{3}}+\left( 2{{{\tilde{K}}}_{P}}\sum\limits_{J}{\tilde{P}_{J}^{2}}-4\sum\limits_{J}{{{{\tilde{A}}}_{J}}} \right)\lambda -2{{\tilde{K}}_{P}}{{\tilde{P}}_{0}}\sum\limits_{J}{{{{\tilde{P}}}_{J}}} = 0 , \label{eq_TissueStressTracelessEquation}
\end{equation}
where ${{\tilde{K}}_{P}} = {K}_{P} / ({K}_{A}{A}_{0})$, ${{\tilde{P}}_{0}} = {P}_{0} / \sqrt{{{A}_{0}}}$, ${{\tilde{A}}_{J}} = {A}_{J} / {A}_{0}$, and ${{\tilde{P}}_{J}} = {P}_{J} / \sqrt{ {A}_{0} }$ are the dimensionless parameters. 
\item Although Eq. \eqref{eq_TissueStressTracelessEquation} does not strictly imply that $\bm{\sigma}^{(\rm tissue)} = \bm{0}$, such condition yields an initial tissue configuration that is relatively close to a zero stress state. To achieve such zero stress state, we let the system evolve through the classical vertex model simulation. We find that this is efficient to decrease the residual global tissue stress. At the end of the simulation, we evaluate whether the stress is small enough; more precisely, we check whether the maximal absolute component of the stress, denoted $|\bm{\sigma}^{(\rm tissue)}|$, is smaller than a threshold value set to $\sigma_0 = 0.01$, i.e. $|\bm{\sigma}^{(\rm tissue)}| < \sigma_0$. If the condition is met, we consider that a stress-free state is obtained; otherwise, we repeat the above steps, until the condition $|\bm{\sigma}^{(\rm tissue)}| < \sigma_0$ is met.
\end{enumerate}
Our protocol was always found to converge.

\subsection{Simple shear simulations} \label{sec:SimpleShearSimulation}

\begin{figure}[h!]
\centering
\includegraphics{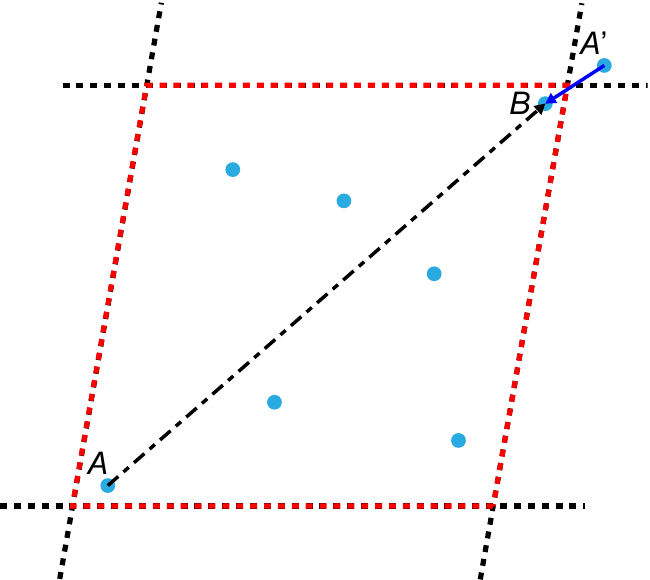}
\caption{\label{fig_LeesEdwardsPeriodicBC} 
Schematic of the Lees--Edwards periodic boundary conditions. 
}
\end{figure}

Our simple shear simulations are performed using the Lees--Edwards periodic boundary conditions (Fig. \ref{fig_LeesEdwardsPeriodicBC} and \cite{Bi2015,Merkel2019}). Note that before shearing, the cell sheet is relaxed to a steady state (see Fig. 1(b) in the main text for examples). In the following we detail our implementation protocol.

\subsubsection{Lees--Edwards periodic boundary conditions}

In the sheared configuration, the calculations of the relative position and distance of vertex (cell) pairs should obey the Lees–Edwards periodic boundary conditions, as shown in Fig. \ref{fig_LeesEdwardsPeriodicBC}. For example, the position vector of the neighboring vertex pair $A$ and $B$, $\bm{r}_{A \rightarrow B}$, cannot be simply measured as $\bm{r}_B - \bm{r}_A$, but instead should be measured as $\bm{r}_B - \bm{r}_{A'}$ with $A'$ being the mirror of vertex $A$. Specifically, denote $\bm{r}_{A \rightarrow B}^{(0)} = \bm{r}_B - \bm{r}_A$, then $\bm{r}_{A \rightarrow B} = (x_{A \rightarrow B}, y_{A \rightarrow B})$ can be calculated as 
\begin{align}
\begin{split}
{{x}_{A\to B}} &= x_{A\to B}^{\left( 0 \right)}-{{L}_{x}}\mathcal{R}\left[ \frac{x_{A\to B}^{\left( 0 \right)}-{{\gamma }_{xy}}y_{A\to B}^{\left( 0 \right)}}{{{L}_{x}}} \right]-{{\gamma }_{xy}}{{L}_{y}}\mathcal{R}\left[ \frac{y_{A\to B}^{\left( 0 \right)}}{{{L}_{y}}} \right], \\ 
{{y}_{A\to B}} &= y_{A\to B}^{\left( 0 \right)}-{{L}_{y}}\mathcal{R}\left[ \frac{y_{A\to B}^{\left( 0 \right)}}{{{L}_{y}}} \right],
\end{split}
\end{align}
where $\mathcal{R}\left[\cdot\right]$ represents the function round to nearest integer and $\gamma_{xy}$ is the simple shear strain. 

\subsubsection{Tissue under a constant simple shear strain}

\paragraph*{Definition} To examine the shear modulus of a cell sheet, we perform simple shear simulations as below:
\begin{enumerate}
    \item First, we apply a small simple shear deformation $\gamma_{xy} = 0.01 \ll 1$ to the cell sheet by mapping all the vertices to new positions, 
\begin{equation}
{{x}_{\text{new}}} = x+{{\gamma }_{xy}}y \quad , \quad {{y}_{\text{new}}} = y.
\end{equation}
    \item Second, the simple shear strain is maintained for a sufficiently long time for the sheared configuration to relax. We keep track the global shear stress $\sigma_{xy}^{(\rm{tissue})}$ during the relaxation and check that the evolution of the global shear falls below a threshold value. The shear modulus of the cell sheet then reads 
\begin{equation}
G_{xy} = \frac{ \langle \sigma _{xy}^{\left( \text{tissue} \right)} \rangle_{\rm A} - \langle \sigma_{xy}^{\left( \text{tissue} \right)} \rangle_{\rm B} }{{{\gamma }_{xy}}},
\end{equation}
where $\langle \sigma _{xy}^{\left( \text{tissue} \right)} \rangle_{\rm B}$ and $\langle \sigma _{xy}^{\left( \text{tissue} \right)} \rangle_{\rm A}$ are the average tissue shear stress before and after simple shear strain, respectively. 
\end{enumerate}
\paragraph*{Verification} We checked that, within the solid regime, the shear modulus is not significantly changed for a range of values $\gamma_{xy} = (0.001,0.10)$ for all considered $\beta$.

\subsubsection{Tissue under quasi-statically increasing simple shear strain}

\paragraph*{Definition} We also perform simple shear simulations by a quasi-static increasing of the amplitude in the simple shear strain $\gamma_{xy}(t)$. The procedure is the following:
\begin{enumerate}
    \item First, the simple shear strain is increased in a step-by-step fashion
\begin{equation}
\gamma _{xy}^{\left( \text{new} \right)}={{\gamma }_{xy}}+\Delta {{\gamma }_{xy}},
\end{equation}
where the incremental shear strain $\Delta\gamma_{xy}$ is small and set to be $\Delta\gamma_{xy} = 0.01$ in our simulations. 

\item Second, after updating the simple shear strain, vertices are mapped to new positions as
\begin{equation}
{{x}_{\text{new}}} = x_{\rm ref} + {{\gamma}_{xy}^{(\rm{new})}}y_{\rm ref} \quad , \quad {{y}_{\text{new}}} = y_{\rm ref},
\end{equation}
where $x_{\rm ref} = x - \gamma_{xy}y$ and $y_{\rm ref} = y$
are the reference coordinates in the reference configuration without shear.

\item Third, the new simple shear strain is kept and the cell sheet is relaxed for a duration $\tau_{\rm relax} = 100$. After relaxation, the global shear stress is computed to construct the stress-strain relation $\sigma_{xy}^{(\rm{tissue})}(\gamma_{xy})$ during such continuous shearing process.
\end{enumerate}

We then repeat the above three steps up to a maximal strain set at $\gamma_{xy} = 3.0$. The tissue-scale shear modulus can be estimated as the derivative at zero shear strain of the shear stress curve with respect to the initial shear strain.  

\paragraph*{Verification} We checked that the relaxation time $\tau_{\rm relax} = 100$ after each step increase is sufficient long to provide stable estimates of the shear modulus and yield stress $\sigma_{\rm yield}$), see Sec. \ref{sec:yieldstress}.

\subsection{Quantification}

\subsubsection{Coordination number $Z$} \label{sec:coordination}

To quantify the coordination number $Z$ represented in Figure 1, we first (1) find clusters of vertices, denoted $C_i$, that are closer to one another than a threshold distance $0.01$ and (2) estimate the mean coordination number 
$Z = \sum_{C_i} N(C_i)$ where $N(C_i)$ is the number of vertices in the cluster $C_i$.

\subsubsection{T1 transition rate}

\begin{figure*}[h!]
 \centering
 \includegraphics[width=16cm]{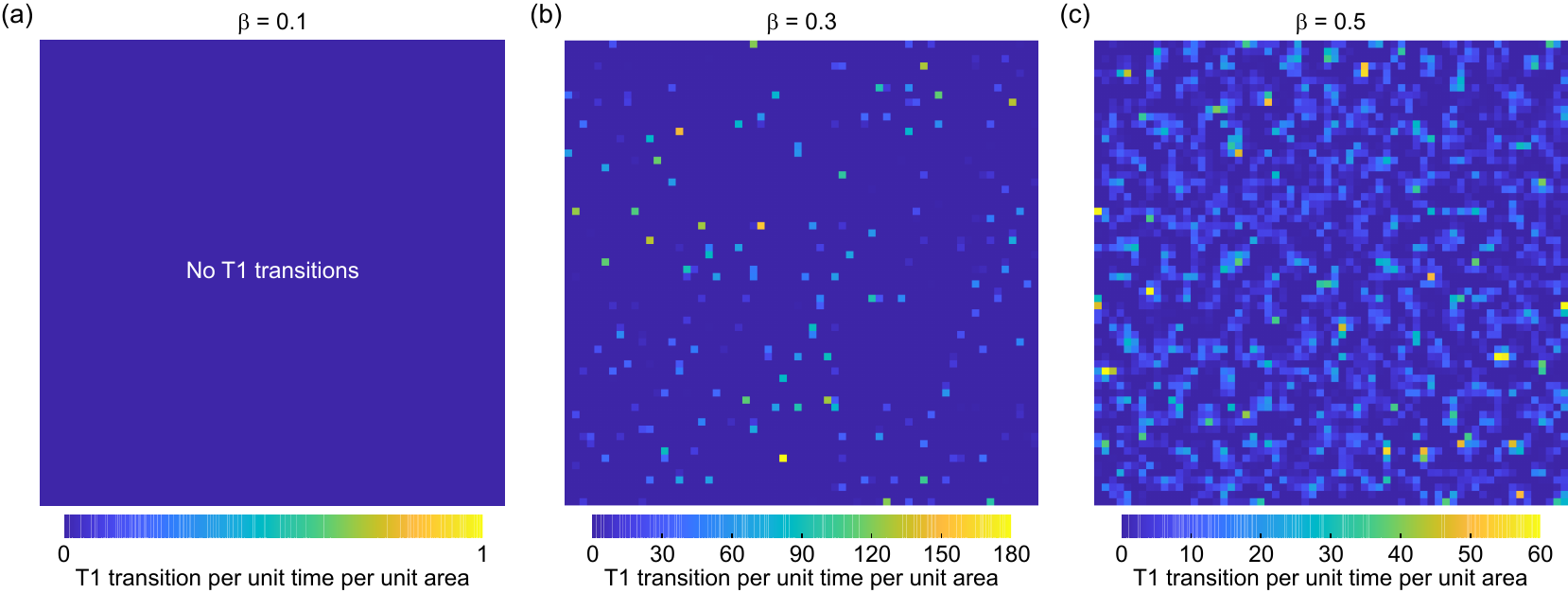}
\caption{\label{fig_T1transition_Locations} 
The hotspot map of T1 transitions (including repeated ones). (a) $\beta = 0.1$. (b) $\beta = 0.3$. (c) $\beta = 0.5$. 
Parameters: $K_P = 0.02$, and $P_0 = 1.0$. 
}
\end{figure*}

\paragraph*{Definition} To quantify the dynamics of cell rearrangements, we examine the rate of T1 transitions after the system arrives at a dynamic steady state. We define the T1 transition rate as, 
\begin{equation} \label{eq:T1definition}
k_{\rm T1} = \frac{N_{\rm T1}}{N_{\rm cells} \tau} , 
\end{equation}
where $N_{\rm T1}$ is the total number of T1 transitions during the observation time $\tau$. 
\paragraph*{Localization of T1 transitions} We also examine the spatial distribution of T1 transitions, as shown in Fig. \ref{fig_T1transition_Locations}. It demonstrates that at the rhombile regime, T1 transitions are highly localized (with hotspots of frequent T1 transitions corresponding to cell-cell rearrangement oscillations) and sub-domains of rhombile patterns without any T1 transitions (see Movie S2). In the fluid regime $\beta = 0.5$, T1 transitions are more uniformly distributed, indicating a completely disordered fluid regime. 

\paragraph*{Method for counting T1 transitions} 
We exclude repeated T1 transitions from the count of the T1 transition rate in the main text Figure 1. To exclude repeated T1 transitions, we keep track of the cells involved in each T1 transition and disregard any new T1 transitions that would involve  the exact same set of cells as previously encountered. The number of T1 transition in Figure 1 was estimated over an observation time window of duration $\tau = 100$ ending with simulations.

\subsubsection{Bond-orientational correlation function}

\paragraph*{Definition} As in the main text, the tissue hexatic order parameter is defined as
\begin{equation}
{{\psi }_{6}}\left( t \right) = \left| \frac{1}{N}\sum\limits_{j=1}^{N}{{{\Psi }_{j}}\left( t \right)} \right| , 
\end{equation}
where ${{\Psi }_{j}} = \sum_{k\in \text{neighbors}}{\exp \left( \text{i}6{{\theta }_{jk}} \right)} / N_j$, with $\theta_{jk} = \arg \left( \bm{r}_k - \bm{r}_j \right)$ \cite{Li2018,Paoluzzi2021}. To quantify the spatial correlation of tissue architecture, we compute the bond-orientational correlation function, 
\begin{equation}
{{g}_{6}}\left( r \right)=\frac{\left\langle {{\Psi }_{i}}\overline{{{\Psi }_{j}}} \right\rangle }{\left\langle {{\Psi }_{i}}\overline{{{\Psi }_{i}}} \right\rangle } , \label{eq_g6}
\end{equation}
where $\overline{\star}$ denotes the conjugate complex number and $r = | \bm{r}_i - \bm{r}_j |$.

\begin{figure}[h!]
\includegraphics[width=14cm]{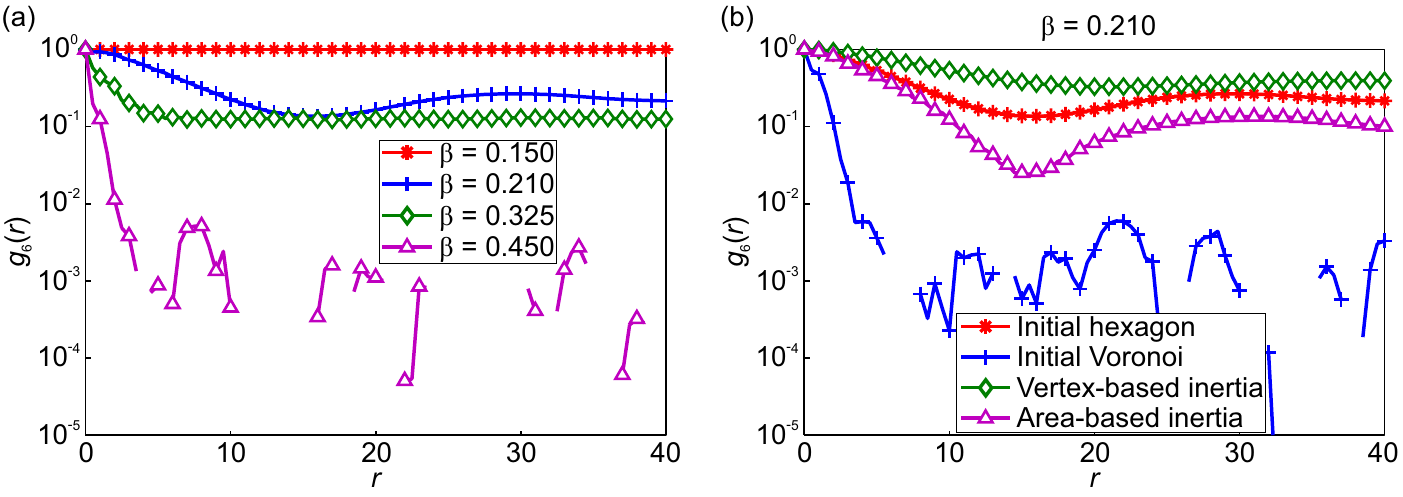} 
\caption{\label{fig_g6} 
The bond-orientational correlation function $g_6(r)$, defined by Eq. \eqref{eq_g6}. (a) $g_6(r)$ at different $\beta$ values, for initially hexagonal pattern. (b) Comparison of $g_6(r)$ for either the initially hexagonal pattern (red: standard $Q$ tensor; green: vertex-based inertia tensor; magenta: area-based inertia tensor) or the initially Voronoi pattern (blue: standard $Q$ tensor), at $\beta = 0.210$. 
Parameters: $K_P = 0.02$, $P_0 = 1$ and $N = 10,000$ cells.
}
\end{figure}

\paragraph*{Results} Figure \ref{fig_g6}(a) shows the bond-orientational correlation function $g_6(r)$ for different values of $\beta$. For the solid regime with isotropic cells, $g_6(r) = 1$; for both the solid regime with anisotropic cells and the rhombile regime, $g_6(r)$ decreases as $r$ is increased and approaches a finite positive value for large $r$, suggesting long-range orientational order when the tissue is initialized according to a near perfect hexagonal pattern.  We confirm such result for the three different definitions of $\bm{Q}$. In contrast, when the tissue is initialized according to a Voronoi pattern, $g_6(r)$ decreases exponentially to zero. In the fluid regime $\beta > \beta_3$ (see main text), $g_6(r)$ decreases exponentially to zero at large $r$.

\subsubsection{Yield shear stress} \label{sec:yieldstress}

\begin{figure*}[h!]
\centering
\includegraphics[width=14cm]{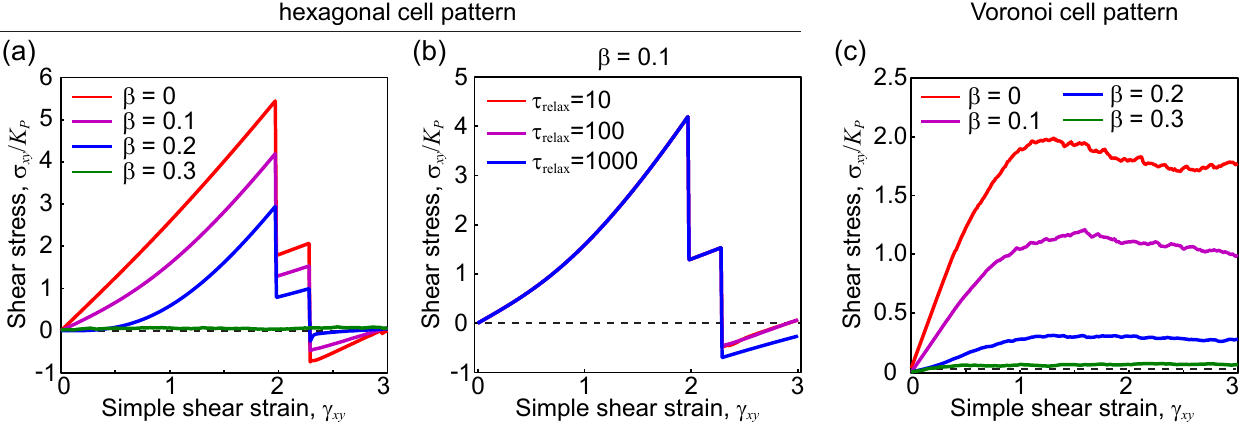}
\caption{\label{fig_YieldStress} 
The yield stress measurement. (a, b) Hexagonal cell pattern; (c) Voronoi cell pattern. 
(a, c) Curves of the shear stress $\sigma_{xy}$ versus the simple shear strain $\gamma_{xy}$ for different cell activities $\beta = 0, \ 0.1, \ 0.2, \ 0.3$. Here the relaxation time $\tau_{\rm relax} = 100$. 
(b) Curves of the shear stress $\sigma_{xy}$ versus the simple shear strain $\gamma_{xy}$ for different relaxation time $\tau_{\rm relax} = 10, \ 100, \ 1000$ at $\beta = 0.1$. 
Parameters: $K_P = 0.02$ and $P_0 = 1.0$. 
}
\end{figure*}

\paragraph*{Definition} In Sec. \ref{sec:SimpleShearSimulation}, we proposed a protocol to implement a quasi-static increase of a simple shear strain. We represent typical stress-strain curves for different values of $\beta$ for tissues either initialized in a near hexagonal state (Fig. \ref{fig_YieldStress}(a)) or according to a Voronoi pattern (Fig. \ref{fig_YieldStress}(c)). We define the yield stress $\sigma_{\rm yield}$ as the maximum of the shear stress $\sigma_{xy}$ curve. The yield stress is represented in the main text Figure 1(e). 

\paragraph*{Discussion} When the initial pattern is near hexagonal (Fig. \ref{fig_YieldStress}(a)), the shear stress sharply drops beyond a critical yield strain. Such a sharp drop of $\sigma_{xy}$ is due to a large number of simultaneous T1 transitions. The curve is more smooth in the case for a tissue initialized according to a Voronoi pattern.

\paragraph*{Verification} We check that the measured shear modulus and yield stress are insensitive to the relaxation time duration of simple shear strain increment in a broad range $10 < \tau_{\rm relax} < 1000$, Fig. \ref{fig_YieldStress}(b). 

\subsection{Simulation check for the passive case}

\begin{figure*}[h!]
\centering
\includegraphics[width=17.5cm]{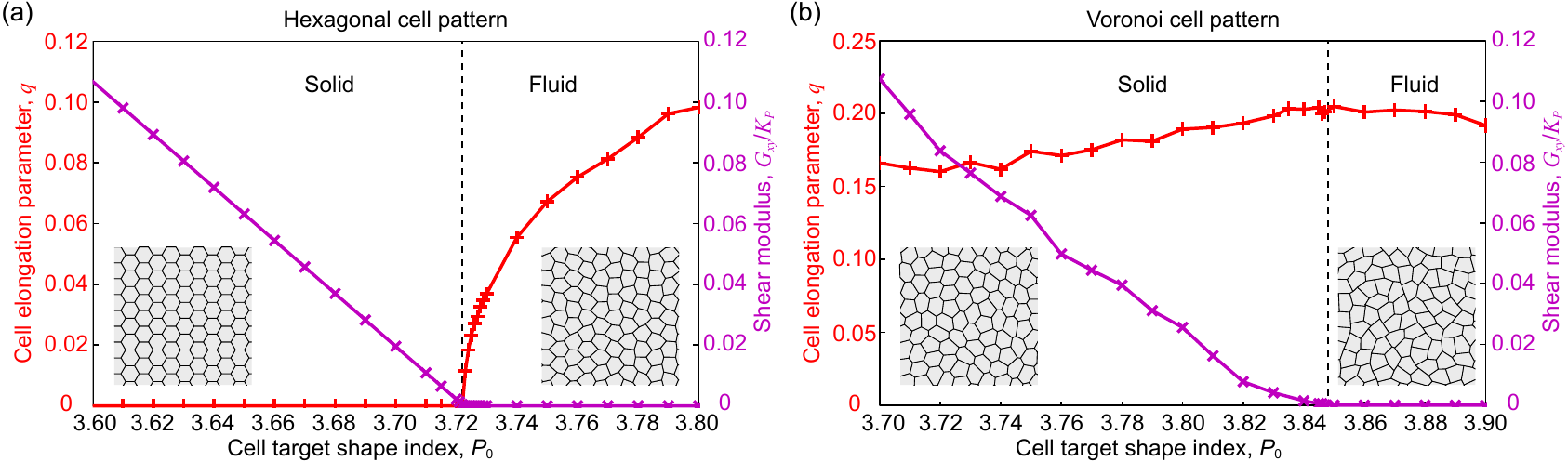}
\caption{\label{fig_RigidityTransition_PassiveVertexModel} 
Simulation of the rigidity transition in the passive vertex model. (a) Hexagonal cell pattern. (b) Voronoi cell pattern. Shown here are the cell elongation parameter $q$ and the (long-time) shear modulus $G_{xy}$ as a function of the cell target shape index $P_0$. 
Parameters: $K_P = 0.02$ and $\beta = 0$. 
}
\end{figure*}

Here we set $\beta = 0$ (passive case) and check that our simulation framework agrees with previous studies \cite{Farhadifar2007,Bi2015,Sussman2018,Merkel2019,Wang2020}. We first check it for an initially hexagonal cell pattern. 
We represent the cell elongation $q$ and shear modulus $G_{xy}$ as function of the target shape index $P_0$ in Fig. \ref{fig_RigidityTransition_PassiveVertexModel}, which clearly indicates a shape and rigidity transition at $P_0^{\ast} \approx 3.723$. 
We also check it for an initially Voronoi cell pattern. Figure \ref{fig_RigidityTransition_PassiveVertexModel} shows that such a cell pattern undergoes a rigidity transition at $P_0^{\ast} \approx 3.846$. 

\clearpage

\section{Stability of the hexagonal pattern} \label{sec:StabilityAnalysis}

\subsection{Shape stability analysis of a single cell}

\begin{figure}[h!]
\ifthenelse{ \cverbose > 0}{}{\includegraphics[width=12.5cm]{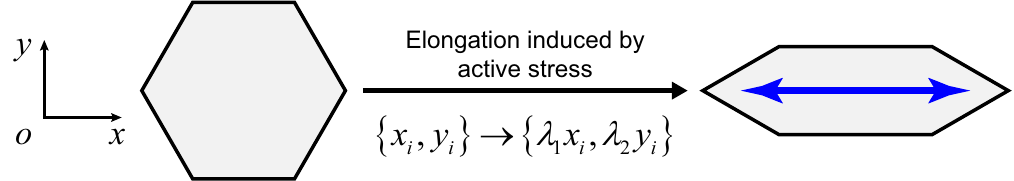}} 
\caption{\label{fig_SingleCellAbalysis} Sketch of cell elongation induced by the cell shape-dependent active stress.}
\end{figure}

Here we consider a regular hexagonal cell with shape-dependent active stress $\bm{\sigma}^{(\rm{act})} = -\beta\bm{Q}$ and ask whether the cell will elongate. To theoretically address such question, we consider a uniform deformation of the cell, as shown in Fig. \ref{fig_SingleCellAbalysis}. The cell deformation can be described by the stretches $\{ \lambda_1, \lambda_2 \}$: $\{ x_i, y_i \} \rightarrow \{ \lambda_1 x_i, \lambda_2 y_i \}$, where $i$ is the index of cell vertices and $\lambda_1$ and $\lambda_2$ are cell stretches along $x$-axis and $y$-axis, respectively. 

\subsubsection{Equilibrium radius of a hexagonal cell without active stress}

We first examine the equilibrium state of a hexagonal cell without active stress. We express the area and perimeter of the cell as $A = \frac{3\sqrt{3}}{2}{{R}^{2}}$ and $P = 6R$ in terms of a quantity $R$, that we call the cell radius. We define the dimensionless equilibrium cell radius  $\tilde{R}_{\rm h} = R_{\rm h} / \sqrt{A_0}$ as the solution of the mechanical equilibrium condition $\bm{\sigma} = \bm{0}$, which reads
\begin{equation}
\frac{9}{4}\tilde{R}_{\text{h}}^{3}+\left( 6{{{\tilde{K}}}_{P}}-\frac{\sqrt{3}}{2} \right){{\tilde{R}}_{\text{h}}}-{{\tilde{K}}_{P}}{{\tilde{P}}_{0}}=0 , \label{eq_EquilibriumEquation_NoActiveStress_dimensionless}
\end{equation}
In the incompressible limit case, i.e., $\tilde{K}_P \ll 1$, the equilibrium cell radius can be easily computed as $\tilde{R}_{\rm h} \simeq {\sqrt{2\sqrt{3}}}/{3} \approx 0.6204$. 

\subsubsection{Equilibrium configuration of a hexagonal cell with active stress}

We now consider the effect of an active stress $\bm{\sigma}^{(\rm{act})} = -\beta\bm{Q}$, in which case the total cell stress reads 
\begin{equation}
\bm{\sigma } = \left[ {{K}_{A}}\left( A-{{A}_{0}} \right)+\frac{1}{2}\frac{{{K}_{P}}P\left( P-{{P}_{0}} \right)}{A} \right]\bm{I}+\left[ \frac{{{K}_{P}}P\left( P-{{P}_{0}} \right)}{A}-\beta  \right]\bm{Q} . \label{eq_Stress_ActiveCase}
\end{equation}
Under an affine deformation:
\begin{equation} \label{eq:lambda1}
\left\{ {{x}_{i}},{{y}_{i}} \right\}\to \left\{ {{\lambda }_{1}}{{x}_{i}},{{\lambda }_{2}}{{y}_{i}} \right\},
\end{equation}
the cell area, perimeter and anisotropy tensor of an initially hexagonal cell read 
\begin{align}
A &= \frac{3\sqrt{3}}{2}R_{\text{h}}^{2}{{\lambda }_{1}}{{\lambda }_{2}} , \label{eq_A_Afine}\\
P &= 2{{R}_{\text{h}}}\left( {{\lambda }_{1}}+\sqrt{\lambda _{1}^{2}+3\lambda _{2}^{2}} \right), \label{eq_P_lambda} \\
\bm{Q}&=\left( \frac{{{\lambda }_{1}}}{\sqrt{\lambda _{1}^{2}+3\lambda _{2}^{2}}}-\frac{1}{2} \right){{\bm{e}}_{x}}\otimes {{\bm{e}}_{x}}-\left( \frac{{{\lambda }_{1}}}{\sqrt{\lambda _{1}^{2}+3\lambda _{2}^{2}}}-\frac{1}{2} \right){{\bm{e}}_{y}}\otimes {{\bm{e}}_{y}} . \label{eq:Q_lambda}
\end{align}

\begin{figure}[t!]
\includegraphics[width=8cm]{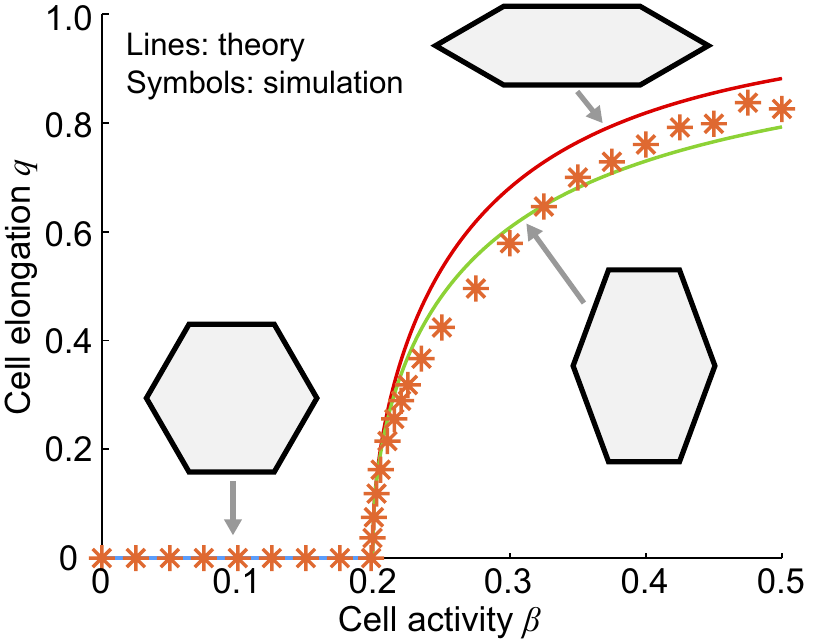} 
\caption{\label{fig_EquilibriumState_q} Comparison of the elongation parameter $q$ of a hexagonal cell versus the cell activity $\beta$ between theoretical predictions (exact solutions given by solving Eqs. \eqref{eq_EquilibriumEquation_Affine_1} and \eqref{eq_EquilibriumEquation_Affine_2}) and single-cell simulations. 
Parameters: $\tilde{K}_P = 0.02$ and $\tilde{P}_0 = 1$ (which results in an estimated transition at $\tilde{\beta}_1 \approx 0.1987$ according to our theoretical prediction Eq. (\ref{eq_beta_cr})).}
\end{figure}

The equilibrium state $\left\{ \lambda_1, \lambda_2 \right\}$ of a cell corresponds to $\bm{\sigma} = \bm{0}$. Substituting the expressions of $A$, $P$ and $\bm{Q}$ into Eq. \eqref{eq_Stress_ActiveCase}, the condition $\bm{\sigma} = \bm{0}$ leads to:
\begin{align}
& \left( \frac{3\sqrt{3}}{2}\tilde{R}_{\text{h}}^{2}{{\lambda }_{1}}{{\lambda }_{2}}-1 \right){{\lambda }_{2}}\sqrt{\lambda _{1}^{2}+3\lambda _{2}^{2}}-\frac{1}{2}{{\lambda }_{2}}\left( 2{{\lambda }_{1}}-\sqrt{\lambda _{1}^{2}+3\lambda _{2}^{2}} \right)\tilde{\beta } \notag \\ 
 & +\frac{4{{{\tilde{K}}}_{P}}}{3\sqrt{3}}\left( {{\lambda }_{1}}+\sqrt{\lambda _{1}^{2}+3\lambda _{2}^{2}} \right)\left[ 2\left( {{\lambda }_{1}}+\sqrt{\lambda _{1}^{2}+3\lambda _{2}^{2}} \right)-\frac{{{{\tilde{P}}}_{0}}}{{{{\tilde{R}}}_{\text{h}}}} \right] = 0 , \label{eq_EquilibriumEquation_Affine_1}
\end{align}
\begin{align}
& \left( \frac{3\sqrt{3}}{2}\tilde{R}_{\text{h}}^{2}{{\lambda }_{1}}{{\lambda }_{2}}-1 \right){{\lambda }_{1}}\sqrt{\lambda _{1}^{2}+3\lambda _{2}^{2}}+\frac{1}{2}{{\lambda }_{1}}\left( 2{{\lambda }_{1}}-\sqrt{\lambda _{1}^{2}+3\lambda _{2}^{2}} \right)\tilde{\beta } \notag  \\ 
 & +\frac{4{{{\tilde{K}}}_{P}}}{\sqrt{3}}{{\lambda }_{2}}\left[ 2\left( {{\lambda }_{1}}+\sqrt{\lambda _{1}^{2}+3\lambda _{2}^{2}} \right)-\frac{{{{\tilde{P}}}_{0}}}{{{{\tilde{R}}}_{\text{h}}}} \right] = 0 . \label{eq_EquilibriumEquation_Affine_2}
\end{align}
where $\tilde{\beta } = {\beta } / ({{{K}_{A}}{{A}_{0}}})$ corresponds to the dimensionless strength of the active stress. Equations \eqref{eq_EquilibriumEquation_Affine_1} and \eqref{eq_EquilibriumEquation_Affine_2} shows that $\{ \lambda_1, \lambda_2 \} = \{ 1, 1 \}$ is always one of the equilibrium state of the single cell system for arbitrary $\beta$. Yet increasing $\beta$, the system undergoes a pitchfork bifurcation and the cell start to elongate, as shown in Fig. \ref{fig_EquilibriumState_q}.

\subsubsection{Stability analysis of a hexagonal cell}

Now we turn to examine the stability of the equilibrium state $\{ \lambda_1^{({\rm eq})}, \lambda_2^{({\rm eq})} \} = \{ 1,1 \}$ (i.e. a regular hexagon at equilibrium radius $R_h$) upon varying $\beta$. The first-order variation of the energy of the single cell system reads 
\begin{equation}
\text{ }\!\!\delta\!\!\text{ }W = \frac{1}{2}\left( \text{ }\!\!\delta\!\!\text{ }A\bm{\sigma }:\bm{\varepsilon }+A\text{ }\!\!\delta\!\!\text{ }\bm{\sigma }:\bm{\varepsilon }+A\bm{\sigma }:\text{ }\!\!\delta\!\!\text{ }\bm{\varepsilon } \right).
\end{equation}
At the equilibrium state $\{ \lambda_1^{({\rm eq})}, \lambda_2^{({\rm eq})} \} = \{ 1,1 \}$, ${{\left. \text{ }\!\!\delta\!\!\text{ }W \right|}_{\left\{ 1,1 \right\}}} = 0$. Further, the second-order variation of the energy reads
\begin{equation}
{{\text{ }\!\!\delta\!\!\text{ }}^{2}}W=\text{ }\!\!\delta\!\!\text{ }A\text{ }\!\!\delta\!\!\text{ }\bm{\sigma }:\bm{\varepsilon }+\text{ }\!\!\delta\!\!\text{ }A\bm{\sigma }:\text{ }\!\!\delta\!\!\text{ }\bm{\varepsilon }+A\text{ }\!\!\delta\!\!\text{ }\bm{\sigma }:\text{ }\!\!\delta\!\!\text{ }\bm{\varepsilon }.
\end{equation}
At the equilibrium state $\{ \lambda_1^{({\rm eq})}, \lambda_2^{({\rm eq})} \} = \{ 1,1 \}$, the latter equation reads
\begin{equation}
{{\left. {{\text{ }\!\!\delta\!\!\text{ }}^{2}}W \right|}_{\left\{ 1,1 \right\}}}={{\left. A \right|}_{\left\{ 1,1 \right\}}}{{\left. \text{ }\!\!\delta\!\!\text{ }\bm{\sigma } \right|}_{\left\{ 1,1 \right\}}}:{{\left. \text{ }\!\!\delta\!\!\text{ }\bm{\varepsilon } \right|}_{\left\{ 1,1 \right\}}} = {{\left. A \right|}_{\left\{ 1,1 \right\}}}\left( {{\varepsilon }_{1}},{{\varepsilon }_{2}} \right){{J}_{\sigma }}{{\left( {{\varepsilon }_{1}},{{\varepsilon }_{2}} \right)}^{\text{T}}} ,  
\end{equation}
where
\begin{equation}
{{J}_{\sigma }}=\left( \begin{matrix}
   \dfrac{3\sqrt{3}}{2}{{K}_{A}}R_{\text{h}}^{2}+{{K}_{P}}\left( 3\sqrt{3}-\dfrac{{{P}_{0}}}{2\sqrt{3}{{R}_{\text{h}}}} \right)-\dfrac{3}{8}\beta  & \dfrac{3\sqrt{3}}{2}{{K}_{A}}R_{\text{h}}^{2}+{{K}_{P}}\left( -3\sqrt{3}+\dfrac{5{{P}_{0}}}{2\sqrt{3}{{R}_{\text{h}}}} \right)+\dfrac{3}{8}\beta   \\
   \dfrac{3\sqrt{3}}{2}{{K}_{A}}R_{\text{h}}^{2}+{{K}_{P}}\left( -3\sqrt{3}+\dfrac{5{{P}_{0}}}{2\sqrt{3}{{R}_{\text{h}}}} \right)+\dfrac{3}{8}\beta  & \dfrac{3\sqrt{3}}{2}{{K}_{A}}R_{\text{h}}^{2}+{{K}_{P}}\left( 3\sqrt{3}-\dfrac{{{P}_{0}}}{2\sqrt{3}{{R}_{\text{h}}}} \right)-\dfrac{3}{8}\beta   \\
\end{matrix} \right) . 
\end{equation}
is the Jacobian matrix at the equilibrium state $\{ \lambda_1, \lambda_2 \} = \{ 1,1 \}$; we further defined the quantities $\varepsilon_1 = \lambda_1 - 1$ and $\varepsilon_2 = \lambda_2 - 1$. The stability of an equilibrium hexagonal cell is equivalent to the positive definiteness of $J_{\sigma}$, 
which results in the set of conditions
\begin{equation}
\left\{ \begin{aligned}
  & 3\sqrt{3}{{K}_{A}}R_{\text{h}}^{2}+\frac{2{{K}_{P}}{{P}_{0}}}{\sqrt{3}{{R}_{\text{h}}}}>0 \\ 
 & {{K}_{P}}\left( 6\sqrt{3}-\frac{\sqrt{3}{{P}_{0}}}{{{R}_{\text{h}}}} \right)-\frac{3}{4}\beta >0 \\ 
\end{aligned} \right.
\end{equation}
The first inequation represents the stability condition for the single cell system without active stress. Therefore, under the shape-dependent active stress $\bm{\sigma} = -\beta\bm{Q}$, the stability condition of a hexagonal cell at rest radius $R_h$ is 
\begin{equation}
\tilde{\beta } < 8\sqrt{3} {{\tilde{K}}_{P}}\left( 1-\frac{{{{\tilde{P}}}_{0}}}{6{{{\tilde{R}}}_{\text{h}}}} \right) \triangleq \tilde{\beta}_1 (\tilde{P}_0) , \label{eq_beta_cr}
\end{equation}
where $\tilde{R}_{\rm h}$ depends on ($\tilde{K}_{P}$, $\tilde{P}_0$), as given by Eq. \eqref{eq_EquilibriumEquation_NoActiveStress_dimensionless}. When $\beta > \beta_1$, the regular hexagonal cell is unstable: under small perturbations, the cell will elongate and cannot recover to a regular hexagonal shape. Our numerical calculations suggest that such cell shape transition belongs to the pitchfork bifurcation class (see Fig. \ref{fig_EquilibriumState_q}). 

Further, in the incompressible limit case, i.e., $\tilde{K}_P \ll 1$, the force balance equation \eqref{eq_EquilibriumEquation_NoActiveStress_dimensionless} gives $\tilde{R}_{\rm h} \simeq {\sqrt{2\sqrt{3}}}/{3} \approx 0.6204$. Then the critical cell activity can be further simplified as: \begin{equation}
{\tilde{\beta }}_1 \simeq {{\tilde{K}}_{P}}\tilde{P}_{0}^{\ast}\left( \tilde{P}_{0}^{\ast}-{{{\tilde{P}}}_{0}} \right) .
\end{equation}
This results is confirmed by our numerical simulations (see main text Fig. 2). 

\subsection{Estimation of the cell elongation beyond the cell-shape instability critical point}

Here we get an analytical prediction for the cell elongation parameter $q$ for $\beta \gtrsim \beta_1$. For simplicity, we consider the incompressible limiting case, i.e., $\tilde{K}_P \ll 1$. In that case, $A \simeq A_0$, $\lambda_2 \simeq 1 / \lambda_1$ (as defined in Eq. \eqref{eq:lambda1}). Based on the expression of $\bm{Q}$ (Eq. \eqref{eq:Q_lambda}), we obtain 
\begin{equation}
q = \sqrt{2\text{tr}\left( {{\bm{Q}}^{2}} \right)} = \left| \frac{2{{\lambda }_{1}}}{\sqrt{\lambda _{1}^{2}+3\lambda _{2}^{2}}}-1 \right| . \label{eq:q_lambda}
\end{equation}
Further assuming $\lambda_1 \geqslant 1$, we get ${{\lambda }_{1}}=\sqrt[4]{{3{{\left( 1+q \right)}^{2}}}/{\left[ \left( 1-q \right)\left( 3+q \right) \right]}}$. We expand $\lambda_1$ and $\lambda_2$ to second order terms in $q$: 
\begin{equation}
{{\lambda }_{1}}(q) = 1+\frac{2}{3}q+\frac{1}{9}{{q}^{2}}+\cdots , \label{eq:lambda1_q}
\end{equation}
\begin{equation}
\lambda _{2}^{{}}(q) = 1-\frac{2}{3}q+\frac{1}{3}{{q}^{2}}+\cdots . \label{eq:lambda2_q}
\end{equation}
Substituting Eqs. \eqref{eq:lambda1_q} and \eqref{eq:lambda2_q} into Eq. \eqref{eq_P_lambda}, we obtain 
\begin{equation}
P(q) \approx P_{0}^{*}+\frac{1}{3}P_{0}^{*}{{q}^{2}}+\cdots  
\end{equation}
Consequently, the anisotropic stress can be expanded to higher order terms of $q$ as, 
\begin{equation}
{{{\tilde{\sigma }}}_{xx}} = \left[ \frac{{{K}_{P}}P\left( P-{{P}_{0}} \right)}{A}-\beta  \right]{{Q}_{xx}} \approx \frac{1}{2}\left[ \frac{{{K}_{P}}P_{0}^{*}\left( P_{0}^{*}-{{P}_{0}} \right)}{A}-\beta  \right]q+\frac{1}{6}\frac{{{K}_{P}}P_{0}^{*}\left( 2P_{0}^{*}-{{P}_{0}} \right)}{A}{{q}^{3}}+\cdots . 
\end{equation}
Combining it with $\tilde{\sigma }={{\partial E}/{\partial \varepsilon }\;}/{\left( 2A \right)}\;$ with $\varepsilon = \lambda_1 - 1 \approx 2 q / 3$, we obtain an effective potential energy of the cell, $E_{\rm eff}(q)$, which reads to fourth order terms in $q$ as: 
\begin{equation}
{{E}_{\text{eff}}}(q) \approx \frac{1}{3}\left[ {{\beta }_{1}}\left( {{P}_{0}} \right)-\beta  \right]{{A}_{0}}{{q}^{2}}+\frac{1}{18}{{K}_{P}}P_{0}^{*}\left( 2P_{0}^{*}-{{P}_{0}} \right){{q}^{4}}+\cdots , \label{eq:EffectivePotential}
\end{equation}
which corresponds to the expression derived in the main text. Minimizing $E_{\rm eff}(q)$ with respect to $q$ yields an approximation of the cell elongation $q_{\min}$: 
\begin{equation}
{{q}_{\min }}=\left\{ \begin{aligned}
  & 0\ \ \ \ \ \ \ \ \ \ \ \ \ \ \ \ \ \ \ \ \ \ \ \ \ \ \ \ \ \ \ \ \ ,\ \ \ \ \ \ \beta <{{\beta }_{1}}\left( {{P}_{0}} \right) \\ 
 & \sqrt{\frac{3\left[ \beta -{{\beta }_{1}}\left( {{P}_{0}} \right) \right]{{A}_{0}}}{{{K}_{P}}P_{0}^{*}\left( 2P_{0}^{*}-{{P}_{0}} \right)}}\ \ \ \ \ \ \ ,\ \ \ \ \ \ \beta >{{\beta }_{1}}\left( {{P}_{0}} \right). \\ 
\end{aligned} \right. \label{eq:q_prediction}
\end{equation}

\begin{figure}[h!]
\includegraphics[width=16cm]{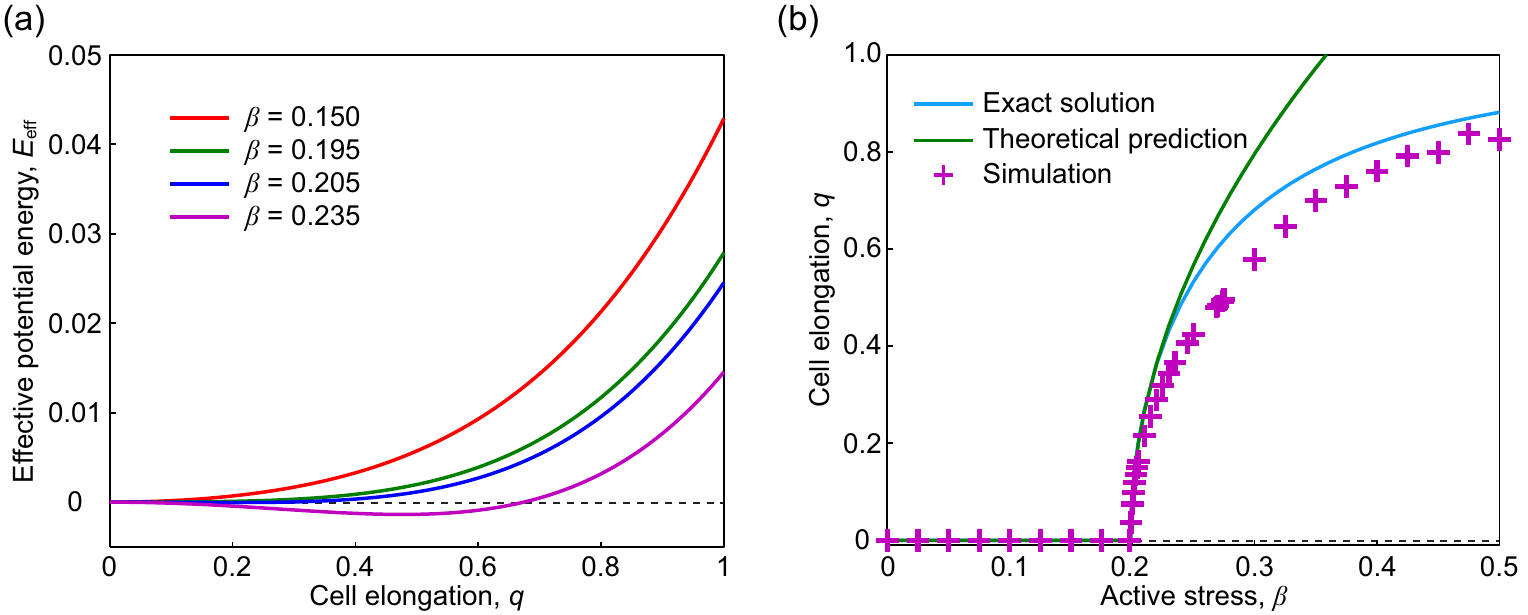}
\caption{\label{fig_q_Prediction} 
Theoretical prediction of the cell elongation parameter $q$. 
(a) The effective potential energy Eq. \eqref{eq:EffectivePotential} as a function of the cell elongation parameter $q$ for different $\beta$ values. 
(b) Comparison of the elongation parameter $q$ of a hexagonal cell versus the cell activity $\beta$ between the exact solutions given by solving Eqs. \eqref{eq_EquilibriumEquation_Affine_1} and \eqref{eq_EquilibriumEquation_Affine_2}, the theoretical predictions given by Eq. \eqref{eq:q_prediction}, and numerical simulations of a single cell system.
Parameters: $\tilde{K}_P = 0.02$ and $\tilde{P}_0 = 1$, which results in $\tilde{\beta}_1 \approx 0.1987$.}
\end{figure}

In Figure \ref{fig_q_Prediction}(a), we represent several $E_{\rm eff}(q)$ curves for several $\beta$ values; this suggest that an instability upon increasing $\beta$ beyond $\beta_1 \approx 0.2$. Figure \ref{fig_q_Prediction}(b) shows the comparison of the theoretical predicted cell elongation $q$ in Eq. \eqref{eq:q_prediction} to the exact cell elongation $q$ obtained by solving Eqs. \eqref{eq_EquilibriumEquation_Affine_1} and \eqref{eq_EquilibriumEquation_Affine_2} and numerical simulations of a single cell (see Sec. \ref{sec:SingleCell_FourCell}). We can see that all match very well for $\beta$ close to the transition point $\beta_1 \approx 0.2$.

\subsection{Simulations of single-cell system and four cell system}
\label{sec:SingleCell_FourCell}

To gain physical insight into the tissue fluidization driven by cell-shape-controlled active stress, we also perform simulations in a single-cell system (Fig. \ref{fig_SingleCellAbalysis}) and a four-cell system (see main text Fig. 1(a), $\beta = 0.150$). In both systems, initial configurations are regular hexagonal patterns with small perturbations applied to each vertex. 

\begin{figure}[h!]
\ifthenelse{ \cverbose > 0}{}{\centering\includegraphics[width=15cm]{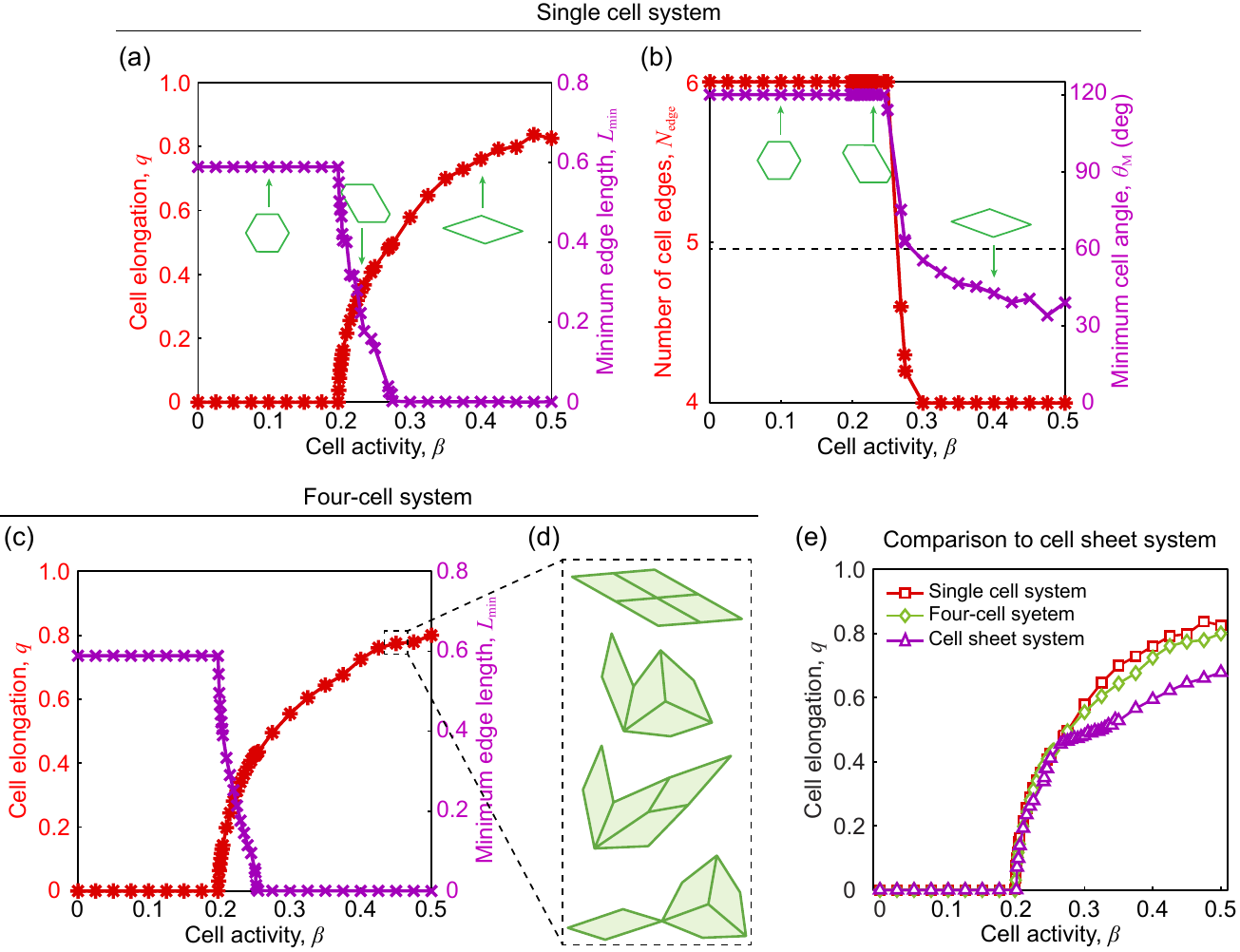}}
\caption{
\label{fig_SingleCellSystem_FourCellSystem_Comparison} 
Morphology transition driven by the active stress in (a,b) the single cell system, (c,d) the four-cell system, and (e) the cell sheet system. 
(a) The cell elongation parameter $q$ and the minimum edge length $L_{\rm min}$ versus the cell activity $\beta$. 
(b) The number of cell edges $N_{\rm edge}$ and the minimum cell angle $\theta_{\rm M}$ versus the cell activity $\beta$. 
(c) The cell elongation parameter $q$ and the minimum edge length $L_{\rm min}$ versus the cell activity $\beta$.  
(d) There exists multiple equilibrium states in the rhombile regime. Shown here are four typical equilibrium states for $\beta = 0.45$. 
(e) Comparison of the cell elongation parameter $q$ in the single cell system, the four-cell system, and the cell sheet system.
Parameters: $K_P = 0.02$, $P_0 = 1$. 
}
\end{figure}

We find that both the single-cell system and the four-cell system exhibit an elongation transition at around $\beta \approx 0.20$ for $P_0 = 1$ (Fig. \ref{fig_SingleCellSystem_FourCellSystem_Comparison}(a,c)), as in larger scale simulations (main text Fig. 1(c)). These two systems also undergo a second transition where some edges vanish. Specifically, the four-cell system undergo such a transition at $\beta \approx 0.25$ (Fig. \ref{fig_SingleCellSystem_FourCellSystem_Comparison}(c)), which corresponds to the end of the solid anisotropic cell regime identified in our larger scale simulations (see also Movie S4 and S5 to visualize these transitions upon increasing $\beta$). 

Further, we measure the minimum cell angle $\theta_M$ in the single-cell system. Interestingly, we find that the single cell system exhibits a perfect rhombile shape with the minimum cell angle $\theta_M \approx 60^{o}$ at $\beta \approx 0.3$. This can explain why we observe the maximum fraction of rhombile regime near $\beta \approx 0.30$. 

In addition, we find that at the higher activity $\beta = 0.4$, there are many metastable states in the four-cell system (Fig. \ref{fig_SingleCellSystem_FourCellSystem_Comparison}(d)). 

Finally, we compare the cell elongation parameter $q$ for simulations of the system-cell system, the four-cell system and the cell sheet system in Fig. \ref{fig_SingleCellSystem_FourCellSystem_Comparison}(e); all agree very well for $\beta$ close to the first transition point.

\clearpage

\section{Stability of the rhombile regime} 

\subsection{The rhombile regime exhibits a finite shear modulus}

\begin{figure*}[h!]
 \centering
 \ifthenelse{ \cverbose > 0}{}{\includegraphics[width=11cm]{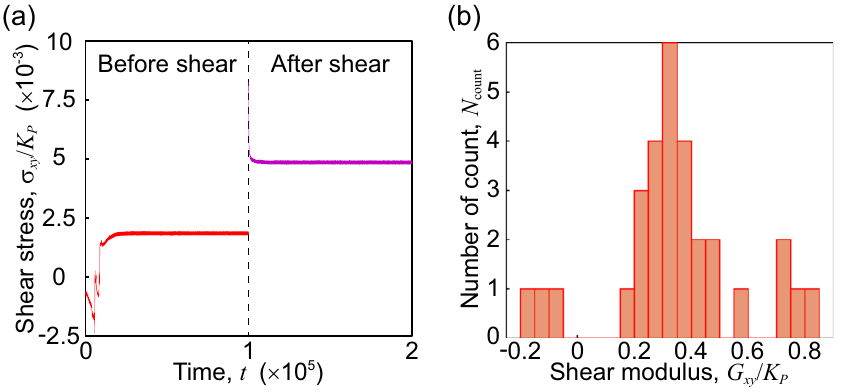} 
 }
\caption{\label{fig_FiniteShearModulus_RhombilePhase} 
The rhombile regime exhibits a finite shear modulus. 
(a) Evolution of the shear stress $\sigma_{xy}$ before and after the application of a small simple shear strain $\gamma_{xy} = 0.01$. Shown here is a typical plot from $n = 30$ independent simulations (initial hexagonal pattern with small perturbations). 
(b) Distribution of the measured shear modulus $G_{xy}$. $n = 30$ independent simulations (initial hexagonal pattern with small perturbations) were performed. 
}
\end{figure*}

We examine more closely the rheological properties of the rhombile regime defined in the main text. In Fig. \ref{fig_FiniteShearModulus_RhombilePhase}(a), we represent a typical evolution curve of the shear stress before and after the application of a small simple shear deformation for $\beta = 0.3$ ($P_0=1$). The sustained increase in the measured shear modulus is consistent with the interpretation that this tissue displays a finite shear modulus. We further perform a statistics in the measured shear modulus for different initial tissue configurations, see Sec. \ref{sec:simulationsetup}; we find that the mean shear modulus is positive, Fig. \ref{fig_FiniteShearModulus_RhombilePhase}(b), which is also consistent with the interpretation that the tissue displays, on average, a finite shear modulus. To confirm that such finite shear modulus is not due to numerical effects, we further investigated the roles of
\begin{itemize}
    \item the T1 transition length threshold $\Delta_{\rm T1}$. We performed simulations with varied T1 transition length threshold $\Delta_{\rm T1}$ and we find that the shear modulus is not sensitive to $\Delta_{\rm T1}$; specifically for $\beta = 0.3$, setting $\Delta_{\rm T1} = 10^{-3}, \ 10^{-2}, \ 10^{-1}$ yielded consistent shear modulus values $G_{xy} / K_P = 0.3244 , \ 0.3000, \ 0.3321$, respectively.
    \item the amplitude of the simple shear strain $\gamma_{xy}$. We performed simulations varying $\gamma_{xy}$ and find that the shear modulus is not sensitive to $\gamma_{xy}$ in a broad range $10^{-5} - 10^{-2}$. Specifically, for $\beta = 0.3$, we measure that $G_{xy} = 0.0069 \pm 0.0046$ for $\gamma_{xy} = 10^{-2}$, $G_{xy} = 0.0068 \pm 0.0046$ for $\gamma_{xy} = 10^{-3}$, $G_{xy} = 0.0069 \pm 0.0047$ for $\gamma_{xy} = 10^{-4}$, $G_{xy} = 0.0071 \pm 0.0047$ for $\gamma_{xy} = 10^{-5}$.
\end{itemize} 
We conclude that the tissue is solid within the identified rhombile regime.

\clearpage

\subsection{Rhombile regime in systems of different sizes} \label{sec:Rhombile_DifferentSystemSize}

\begin{figure*}[h!]
\centering
\includegraphics[width=17.5cm]{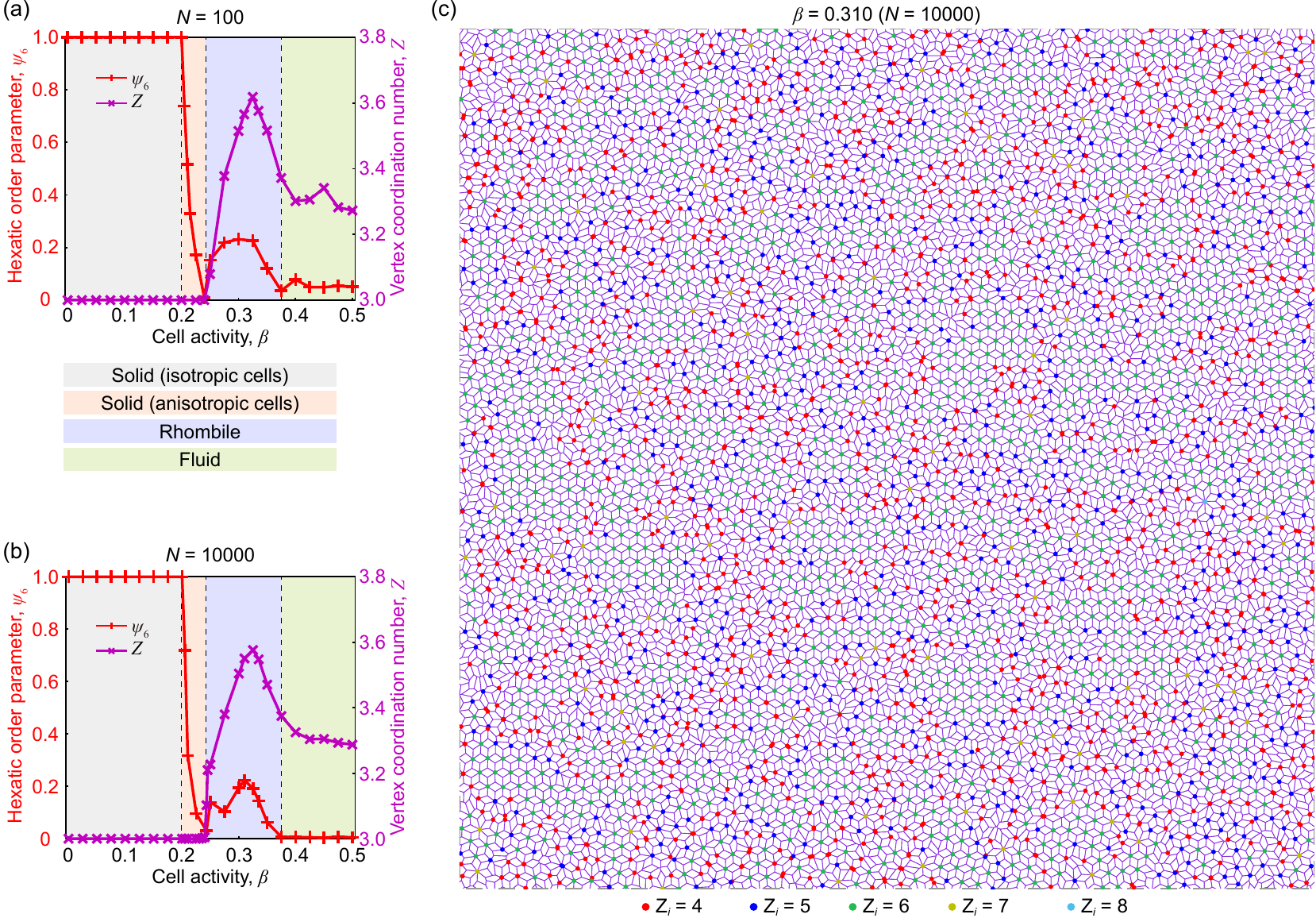}
\caption{\label{fig_SystemSize} 
Active-stress driven tissue fluidity transition in cell sheets of different sizes. 
(a, b) The hexatic order parameter $\psi_6$ and average vertex coordination number $Z$ versus cell activity $\beta$: (a) $N = 100$; (b) $N = 10,000$. 
(c) The rhombile regime ($\beta = 0.31$) in a large cell sheet system ($N = 10,000$). 
Parameters: $K_P = 0.02$, and $P_0 = 1.0$. 
}
\end{figure*}

To explore the effect of system size on the active stress driven tissue fluidity transition in cell sheets, here we vary the system size from a small one ($N = 100$) to a large one ($N = 10,000$). 
We find similar tissue fluidity transition regimes: as the cell activity $\beta$ is increased from small values to large values, the tissue transitions from the solid regime (isotropic cells), the solid regime (anisotropic cells), the rhombile regime, to finally the fluid regime, as shown in Fig. \ref{fig_SystemSize}(a,b). Moreover, the critical $\beta$ values between these regimes for a large system ($N = 10,000$) reads: $\beta_1 \approx 0.20$, $\beta_2 \approx 0.24$, and $\beta_3 \approx 0.37$. These critical $\beta$ values are close to those presented in the main text. 
Finally, we present the rhombile regime in a large system ($N = 10,000$) in Fig. \ref{fig_SystemSize}(c).

\clearpage

\subsection{Stability of a perfect rhombile pattern}

\begin{figure*}[h!]
 \centering
 \ifthenelse{ \cverbose > 0}{}{\includegraphics[width=17.5cm]{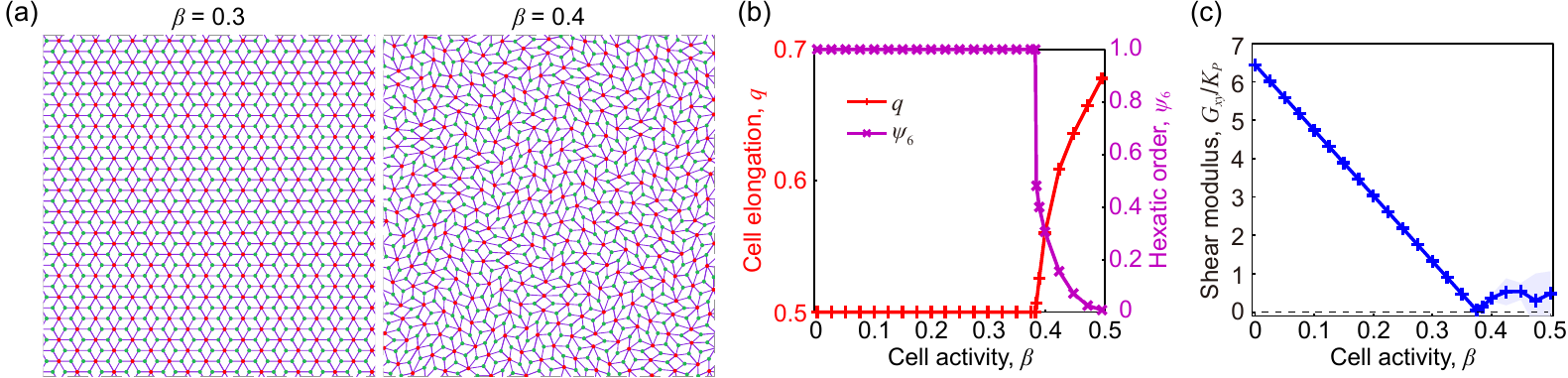} 
 }
\caption{\label{fig_rhombile tiling} 
Stability of the rhombile tiling pattern. Simulations are started from a rhombile tiling with small perturbations $\sigma =0.01$ in the position of each vertex.
(a) Morphology of a cell sheet at different $\beta$ values. 
(b) The cell elongation $q$ and hexatic order $\psi_6$ versus cell activity $\beta$. 
(c) The long-time shear modulus $G_{xy}$ versus cell activity $\beta$. 
Parameters: $K_P = 0.02$ and $P_0 = 1.0$. 
}
\end{figure*}

Here we  examine the stability of a global rhombile tiling pattern, as shown in Fig. \ref{fig_rhombile tiling}(a), where cells are of identical diamond shape and display both 3-way and 6-way vertices. Here, the 6-fold vertices, being considered as a single vertex, cannot split into two separate 3-way vertices. 

Our simulations show that the rhombile tiling pattern becomes unstable as the activity parameter $\beta$ is increased beyond the threshold value $\beta_{\rm cr,3} \approx 0.38$, see Fig. \ref{fig_rhombile tiling}(a,b). The value $\beta_{\rm cr,3} \approx 0.38$ matches the value mentioned in the main text as delineating the rhombile regime from the disordered fluid regime. 

We further estimated the shear modulus of the rhombile tiling pattern through numerical simulation, see Fig. \ref{fig_rhombile tiling}(c) and find that the rhombile tiling pattern is of finite shear modulus for $\beta < \beta_{\rm cr,3}$. The presence of nucleated rhombile crystals may explain why a residual shear modulus can be observed within the rhombile regime. 

\clearpage

\section{Alternative cell shape descriptor} \label{sec:robustness}


\subsection{Vertex-based inertia tensor}

\paragraph*{Definition} Here we first consider an alternative cell shape description to $\bm{Q}$ defined in terms of a vertex-based inertia tensor 
\begin{align} \label{eq:Mdefinition}
\bm{M}_v = \sum_{i}{ {{\bm{\rho }}_{i}}\otimes {{\bm{\rho }}_{i}} }, 
\end{align}
where we have introduced the relative vertex position vectors
$\bm{\rho}_i = \bm{r}_i - \bm{r}_{\rm C}$
with respect to the cell center $\bm{r}_{\rm C} = \sum_k \bm{r}_k / N$. By analogy to $\bm{Q}$ defined in main text, which is a deviatoric tensor, we define the following vertex-based inertia $\bm{Q}$-tensor: 
\begin{equation}
\bm{Q}_{\rm inertia} =\frac{1}{2} \left[ \frac{1}{\text{tr}\left( \bm{M}_v \right)}\bm{M}_v-\frac{1}{2}\bm{I} \right]. \label{eq_Q_definition_Inertia}
\end{equation}
\vskip0.1cm

\begin{figure*}[b!]
 \centering
 \ifthenelse{ \cverbose > 0}{}{\includegraphics[width=15.5cm]{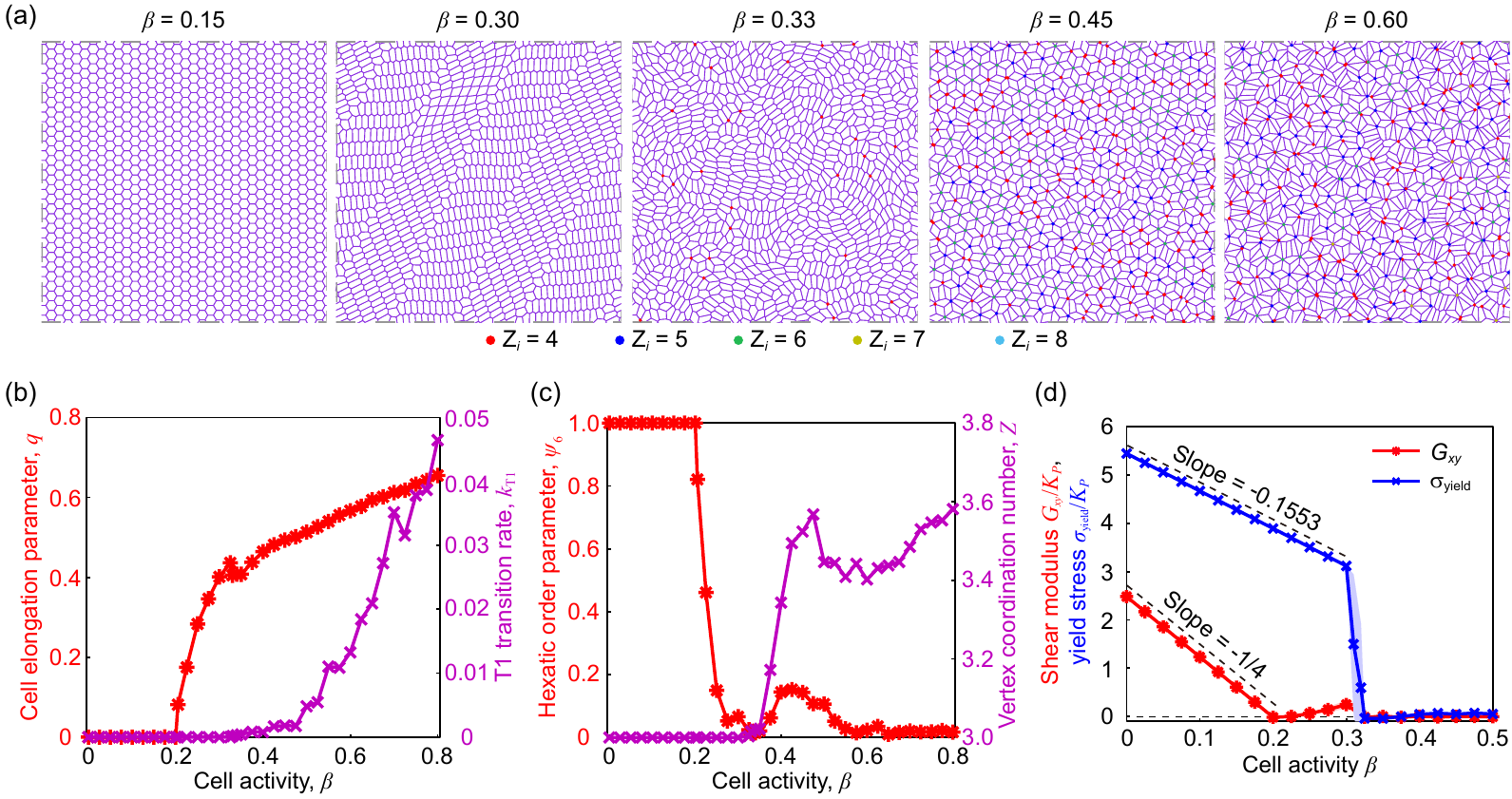} 
 }
 \caption{\label{fig_Q_inertia} 
 Solid-to-fluid transition driven by the active stress $\bm{\sigma}^{\rm (act)} = - \beta \bm{Q}_{\rm inertia}$, where $\bm{Q}_{\rm inertia}$ is defined by Eq. \eqref{eq_Q_definition_Inertia}. 
(a) Morphology of a cell sheet at different cell activities; we mark vertices with coordination number larger than 3, see Sec. \ref{sec:coordination}.
(b) The cell elongation parameter $q$ and the T1 transition rate $k_{\rm T1}$ versus the cell activity $\beta$. 
(c) The hexatic order parameter $\psi_6$ and the average vertex coordination number $Z$ versus the cell activity $\beta$. 
(d) The (long-time) shear modulus $G_{xy}$ and the yield shear stress $\sigma_{\rm yield}$ versus the cell activity $\beta$. 
Parameters: $K_P = 0.02$ and $P_0 = 1.0$. 
}
\end{figure*}

\paragraph*{Results} We find that the hexagonal pattern is destabilized at a critical coupling value $\beta_c \approx 0.2$, see Fig. \ref{fig_Q_inertia}(a,b) in favor of a solid regime with anisotropic cells in the range $\beta \in (0.2,0.35)$. Within the latter range of activity, we observe buckling along tissue-scale stripes, reminiscent of the chevron pattern within smectic A liquid crystal \cite{Taylor}. Such chevron pattern is of finite shear modulus, Fig. \ref{fig_Q_inertia}(d). At a second critical value $\beta_c \approx 0.35$, the vertex coordinate number $Z$ increases, see Fig. \ref{fig_Q_inertia}(c), which marks the end of the solid regime with anisotropic cells. A rhombile crystal domains, with characteristic high proportion of periodically arranged six-fold vertices, are observed within an intermediate range of the coupling value $\beta$; here $\beta \in (0.35,0.58)$; in the latter range, both the vertex coordinate number $Z$ and hexatic order display a peak. For larger value of the coupling $\beta > 0.58$, spontaneous flows take place. 

\paragraph*{Discussion} Overall, the results for the vertex-based inertia cell shape tensor are similar to the one presented in the main text. Following Ref. \cite{Nestor-Bergmann_MMB_2018}, one can show that the inertia tensor $\bm{M}_v$ displays the same principal orientations as the cell nematic shape tensor $\bm{Q}$ defined in the main text; yet the associated eigenvalues for these two tensors are, in general, different. 

\subsection{Area-based inertia tensor}

\paragraph*{Definition} We next consider an area-based inertia tensor $\bm{M}_a$, defined as, 
\begin{equation}
\bm{M}_a = \int{\left( {{\left| \bm{r} \right|}^{2}}\bm{I}-\bm{r}\otimes \bm{r} \right){{\text{d}}^{2}}\bm{r}} , 
\end{equation}
where $\bm{r}$ is the position vector relative to the cell centroid. Similarly, we define the following $\bm{Q}$-tensor: 
\begin{equation}
\bm{Q}_{\rm inertia} = \frac{5}{8} \left[ \frac{1}{2}\bm{I}-\frac{1}{\text{tr}\left( \bm{M}_a \right)}\bm{M}_a  \right]. \label{eq_Q_definition_Inertia_2}
\end{equation}

\begin{figure*}[b!]
 \centering
 \ifthenelse{ \cverbose > 0}{}{\includegraphics[width=15.5cm]{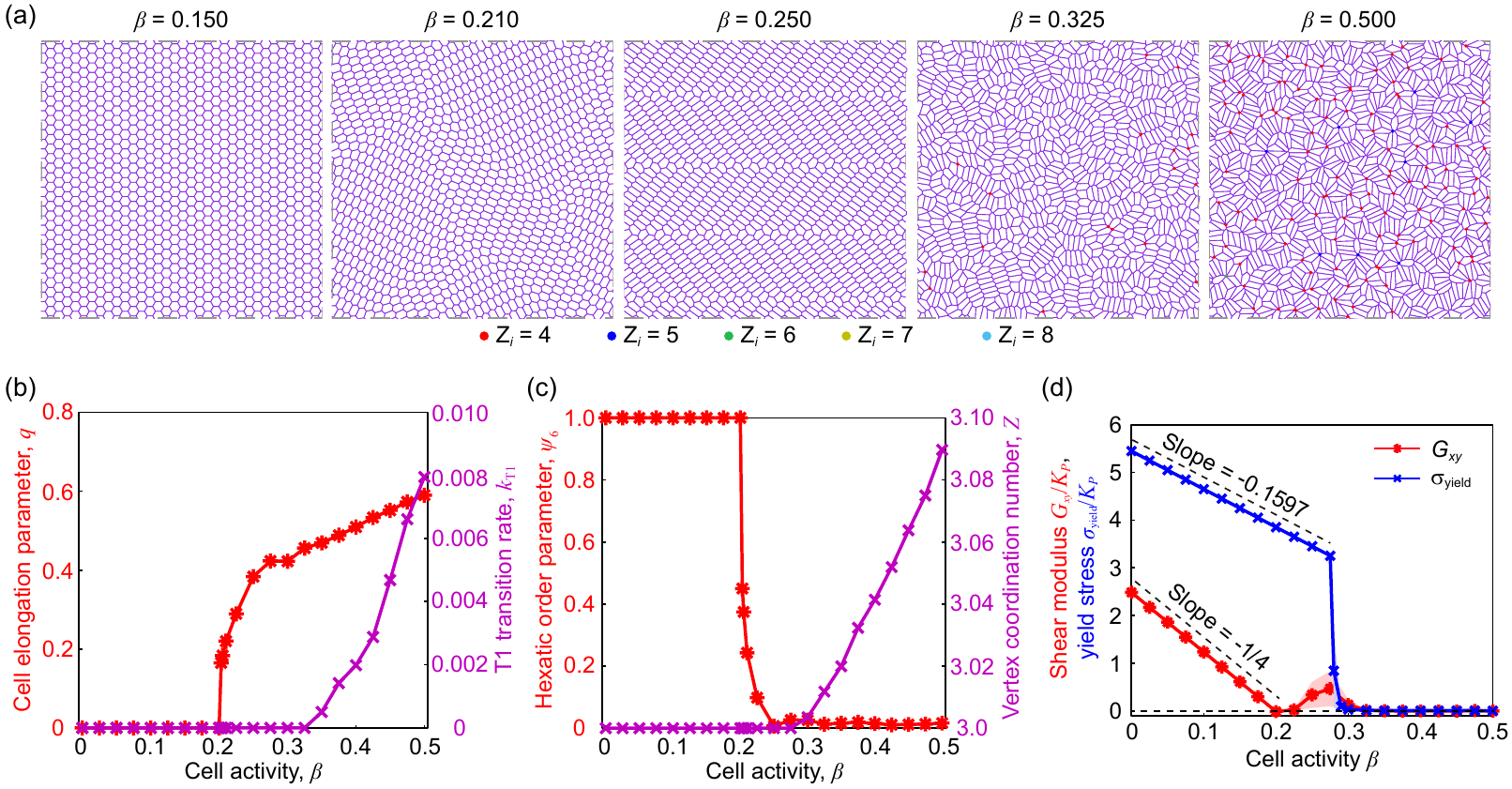} 
 }
 \caption{\label{fig_Q_inertia_2} 
 Solid-to-fluid transition driven by the active stress $\bm{\sigma}^{\rm (act)} = - \beta \bm{Q}_{\rm inertia}$, where $\bm{Q}_{\rm inertia}$ is defined by Eq. \eqref{eq_Q_definition_Inertia_2}. 
(a) Morphology of a cell sheet at different cell activities; we mark vertices with a coordination number larger than 3.
(b) The cell elongation parameter $q$ and the T1 transition rate $k_{\rm T1}$ versus the cell activity $\beta$. 
(c) The hexatic order parameter $\psi_6$ and the average vertex coordination number $Z$ versus the cell activity $\beta$. 
Parameters: $K_P = 0.02$ and $P_0 = 1.0$. 
}
\end{figure*} 

\vskip0.25cm

\paragraph*{Results - similarities with other schemes} We find that the hexagonal pattern is destabilized at a critical coupling value $\beta_c \approx 0.2$, see Fig. \ref{fig_Q_inertia_2}(a,b) in favor of a solid regime solid with anisotropic cells. Within the range $\beta \in (0.2,0.27)$, we also observe buckling along tissue-scale stripes (Fig. \ref{fig_Q_inertia_2}(a)) that are similar to those obtained with the vertex-based inertia tensor (Eq. \eqref{eq_Q_definition_Inertia}); the corresponding tissue state display a finite shear modulus (Fig. \ref{fig_Q_inertia_2}(d)).  At a second critical value $\beta_2 \approx 0.27$, the vertex coordinate number $Z$ increases, see Fig. \ref{fig_Q_inertia_2}(c), which marks the end of the anisotropic solid regime. 
For larger value of the activity parameter $\beta > \beta_3 = 0.32$, spontaneous flows emerge. 

\paragraph*{Results - differences with other schemes} In the intermediate activity $(\beta_2,\beta_3)$, we observe some higher order vertices, but less than in the other schemes for $Q$. We do not observe well-defined rhombile crystal domains, Fig. \ref{fig_Q_inertia_2}(a,c). 
\vskip0.25cm
\paragraph*{Discussion} With the area-based inertia definition, the rhombile crystal pattern is not observed as $\beta$ is increased. We interpret that through a single cell shape analysis, which leads to less of shrinkage at cell edges in the case of the area-based inertia tensor. At the tissue level, this is reflected by a much smaller T1 transition rate $k_{\rm T1}$ (Fig. \ref{fig_Q_inertia_2}(b)) than for the standard inertia tensor $\bm{Q}$ defined on cell edges and Fig. 1(c)) or the vertex-based inertia tensor (Eq. \eqref{eq_Q_definition_Inertia} and Fig. \ref{fig_Q_inertia}(b)).

\clearpage

\section{Cell shape texture and topological defects}

\subsection{Detection of topological defects}

\begin{figure}[b!]
\ifthenelse{ \cverbose > 0}{}{\includegraphics[width=16cm]{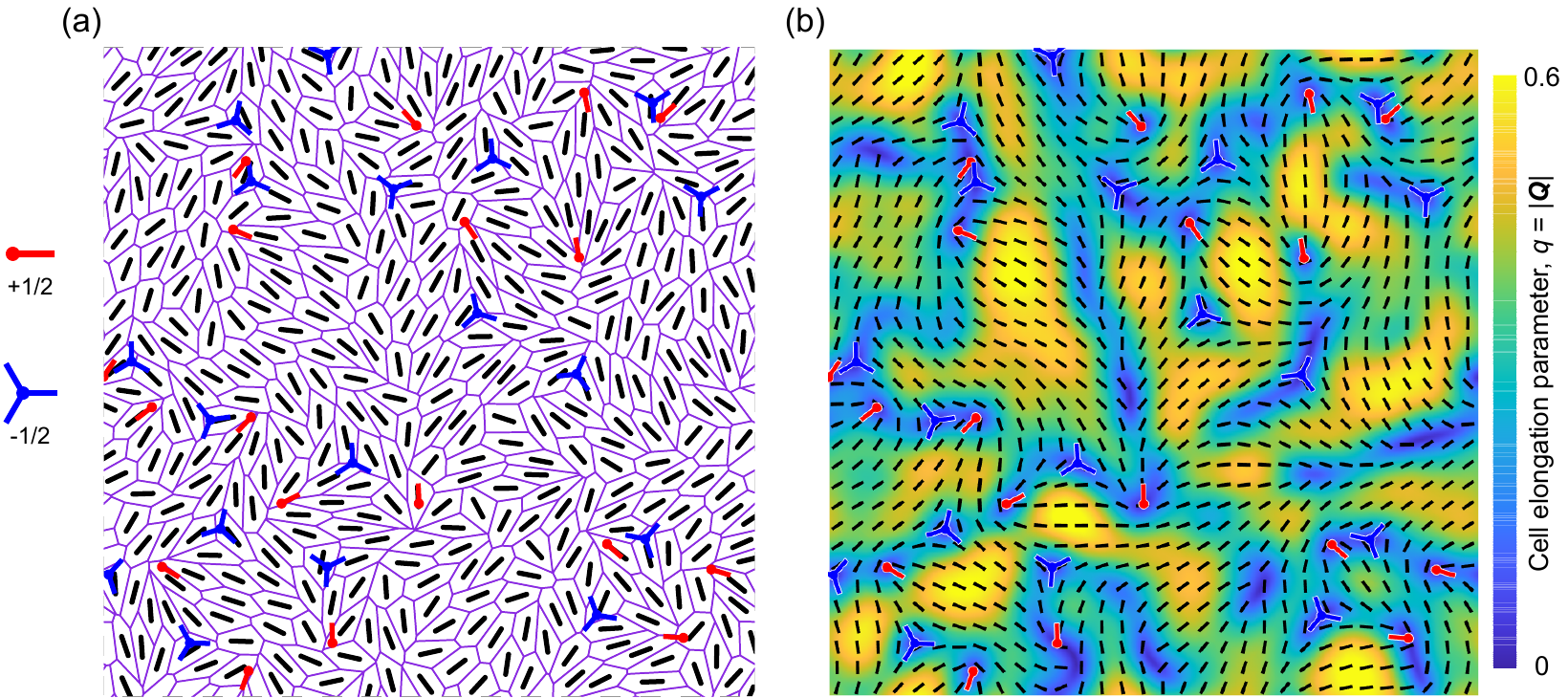}}
\caption{\label{fig_DefectDetectionScheme} Topological defect detection scheme. (a) The original cell sheet morphology superimposed with the detected topological defects (red and blue markers). The black lines indicate cell orientations, given by the cell shape nematic tensors $\bm{Q}_J$. (b) The coarse-grained nematic tensor field superimposed with the detected topological defects, where the color code refers to the cell elongation parameter $q = | \bm{Q} |$ and the black lines indicate local nematic orientations given by $\bm{Q}$. }
\end{figure}

In presence of cell active stress, for $\beta > \beta_1$, a cell sheet exhibits elongated cell shape and coordinated orientations in local areas that are conflicting at singular points, called topological defects (see Fig. \ref{fig_DefectDetectionScheme}(a)). 

To detect the locations and orientations of such topological defects, we first compute the cell geometric centers $\bm{r}_J = \sum_{i \in neighbor} \bm{r}_i / N_J$ with $N_J$ being the number of vertices belonging to the $J$-th cell, and the nematic tensors $\bm{Q}_J$ of cells. Based on $\bm{r}_J$ and $\bm{Q}_J$, we next construct a coarse-grained nematic field $\bm{Q}(\bm{r})$ on a regular grid, 
\begin{equation}
\bm{Q}\left( \bm{r} \right)=\frac{\sum\limits_{\left| \bm{r}-{{\bm{r}}_{J}} \right|<{{r}_{\text{cut-off}}}}{w\left( \bm{r}-{{\bm{r}}_{J}} \right)\bm{Q}_J}}{\sum\limits_{\left| \bm{r}-{{\bm{r}}_{J}} \right|<{{r}_{\text{cut-off}}}}{w\left( \bm{r}-{{\bm{r}}_{J}} \right)}} , 
\end{equation}
where $w(\bm{r}-\bm{r}_J)$ is the weight function and $r_{\rm cut-off} = 3\sigma$ is a cut-off length. Here we set the weight function as a Gaussian function, 
\begin{equation}
w\left( \bm{r}-{{\bm{r}}_{J}} \right)=\dfrac{1}{\sqrt{2\text{ }\!\!\pi\!\!\text{ }}\sigma }\exp \left( -\dfrac{1}{2}\dfrac{{{\left| \bm{r}-{{\bm{r}}_{J}} \right|}^{2}}}{{{\sigma }^{2}}} \right) , \label{eq:WeightFunction}
\end{equation}
where we set the kernel size $\sigma$ at $\sigma = 0.75$ cell length. We set the cut-off length at $r_{\rm cut-off} = 3\sigma = 2.25$ cell length. 

After obtaining the coarse-grained nematic field $\bm{Q}(\bm{r})$, we detect the location of topological defects based on the calculation of winding number \cite{Saw_Nature_2017,Kawaguchi_Nature_2017}. The orientation of topological defects is extracted following the method presented in \cite{Vromans_SoftMatter_2016}; for a topological charge $k$ ($k = \pm 1/2$), the orientation is given by
\begin{equation}
\psi = \frac{k}{1-k}\arctan 2\left[ \left\langle {\rm sgn} \left( k \right)\frac{\partial {{Q}_{xy}}}{\partial x}-\frac{\partial {{Q}_{xx}}}{\partial y} \right\rangle ,\left\langle \frac{\partial {{Q}_{xx}}}{\partial x} + {\rm sgn} \left( k \right)\frac{\partial {{Q}_{xy}}}{\partial y} \right\rangle  \right] , 
\end{equation}
where $\langle \cdot \rangle$ denotes an average along the shortest available loop
enclosing the defect core; $\rm sgn (\cdot)$ is the sign function; $\arctan 2 (y, x)$ is the 2-argument arctangent function.

To check the reliability of such detection scheme, we superimpose the detected topological defects on the original cell sheet morphology and coarse-grained nematic field, as shown in Fig. \ref{fig_DefectDetectionScheme}. We can see that the detected topological defects fit well with both the original cell sheet morphology and the coarse-grained nematic field. We mainly observe two kinds of topological defects, i.e., comet-like $+1/2$ topological defects and trefoil-like $-1/2$ topological defects. 

\subsection{Stress and flow fields near topological defects}

 We examine the average stress and flow fields near $\pm 1/2$ topological defects in our vertex model implementation. The resulting fields are consistent with the predictions of an incompressible active nematic continuum theory \cite{Saw_Nature_2017}.
\vskip0.2cm
\textit{Method} -- We first examine the average isotropic stress ($\sigma_{\rm iso} = (\sigma_{xx} + \sigma_{yy}) / 2$) field. We first define a local coordinate system of a topological defect: its origin locates at the defect core, with $x$-axis along the defect orientation. The average isotropic stress $\hat{\sigma }_{\text{iso}}$ at position $\bm{x}$ (local coordinate system) is estimated through the following expression
\begin{equation}
{\hat{\sigma }_{\text{iso}}}\left( \bm{x} \right)=\frac{\sum\limits_{\left| \bm{x}-{{{\bm{\bar{r}}}}_{J,\alpha}} \right|<{{r}_{\text{cut-off}}}}{w \left( \bm{x}-{{{\bm{\bar{r}}}}_{J,\alpha}} \right)\sigma _{\text{iso}}^{\left( J \right)}}}{\sum\limits_{\left| \bm{x}-{{{\bm{\bar{r}}}}_{J,\alpha}} \right|<{{r}_{\text{cut-off}}}}{w \left( \bm{x}-{{{\bm{\bar{r}}}}_{J,\alpha}} \right)}} .  
\end{equation}
Here $\sigma_{\rm iso}^{(J)}$ is the isotropic stress of the $J$-th cell; $w(\cdot)$ is a Gaussian weight function (Eq. \eqref{eq:WeightFunction}); $\bar{\bm{r}}_{J,\alpha}$ represents the position vector of the geometric centre of the $J$-th cell relative to the $\alpha$-th topological defect, and can be calculated according to the following expression:
\begin{equation}
{{\bm{\bar{r}}}_{J,\alpha }}=\mathcal{R}\left[ \left( {{\bm{r}}_{J}}-{{\bm{x}}_{\alpha }} \right); -{{\theta }_{\alpha }} \right] , 
\end{equation}
where $\theta_{\alpha}$ refers to the orientation of the $\alpha$-th topological defect; $\mathcal{R}[\bm{r} ; \theta]$ represents a rotation of the vector $\bm{r}$ by an angle $\theta$. 
Similarly, the average flow velocity $\hat{\bm{v}}(\bm{x})$ at position $\bm{x}$ (local coordinate system) is estimated as
\begin{equation}
\bm{\hat{v}}\left( \bm{x} \right)=\frac{\sum\limits_{\left| \bm{x}-{{{\bm{\bar{r}}}}_{J,\alpha }} \right|<{{r}_{\text{cut-off}}}}{\omega \left( \bm{x}-{{{\bm{\bar{r}}}}_{J,\alpha }} \right)\mathcal{R}\left[ {{\bm{v}}_{J}};-{{\theta }_{\alpha }} \right]}}{\sum\limits_{\left| \bm{x}-{{{\bm{\bar{r}}}}_{J,\alpha }} \right|<{{r}_{\text{cut-off}}}}{\omega \left( \bm{x}-{{{\bm{\bar{r}}}}_{J,\alpha }} \right)}} , 
\end{equation}
where $\bm{v}_J = \mathrm{d}\bm{r}_J / \mathrm{d}t$ is the motion velocity of the $J$-th cell.

\begin{figure}[h!]
\includegraphics[width=17.5cm]{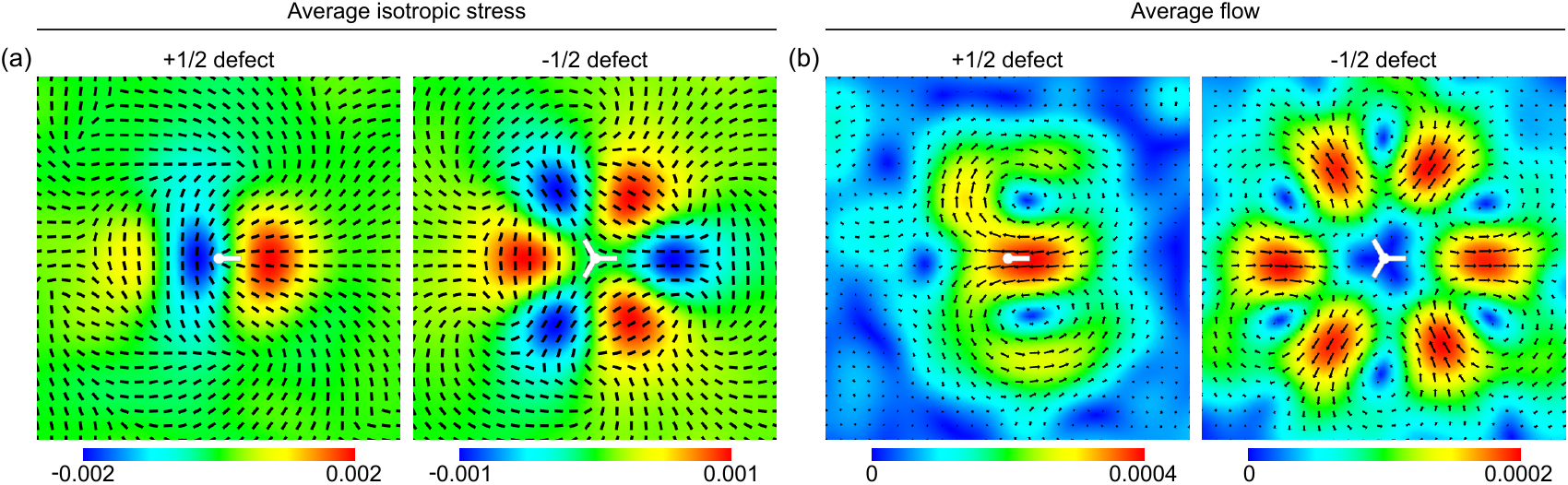}
\caption{\label{fig_Fields_TopologicalDefects} The isotropic stress field (a) and flow field (b) near half-integer topological defects, averaged over thousands of topological defects ($n_{\pm 1/2} = 9,844$). In (a), the color code refers to the local stress fluctuation $\sigma_{\rm iso} - \langle \sigma_{\rm iso} \rangle$ and the black lines indicate local cell orientations. In (b), the color code refers to the velocity magnitude $| \bm{v} |$ and black arrows indicate local flow velocity vectors. Domain size: $L = 10$. Parameters: $K_P = 0.02$, $P_0 = 1$, and $\beta = 0.4$. }
\end{figure}

\textit{Result} -- Figure \ref{fig_Fields_TopologicalDefects} shows the coarse-grained average isotropic stress and flow fields near $\pm 1/2$ topological defects, for $\beta = 0.4 > 0$ (extensile active nematic case). We observe that  both the stress and flow fields are consistent with previous experimental measurements (including MDCK cell monolayer system \cite{Saw_Nature_2017,Balasubramaniam2021}, human bronchial epithelial cell monolayer system \cite{Blanch-Mercader_PRL_2018} and neural progenitor cell monolayer system \cite{Kawaguchi_Nature_2017}) and predictions of a continuum active gel theory \cite{Giomi2015,Saw_Nature_2017}.

\subsection{Cell-cell rearrangement field near topological defects}
In the main text Figure 3, we examine the cell-cell rearrangement rate field near topological defects. We define the cell rearrangement rate as the number of T1 transitions per unit time per unit area, estimated according to the following formula 
\begin{equation}
\chi \left( \bm{x} \right) = \frac{\sum\limits_{\left| \bm{x}-{{{\bm{\bar{r}}}}_{T,\alpha }} \right|<{{r}_{\text{cut-off}}}}{\delta \left( \bm{x}-{{{\bm{\bar{r}}}}_{T,\alpha }} \right)}}{\text{ }\!\!\pi\!\!\text{ }r_{\text{cut-off}}^{2}\tau } , 
\end{equation}
where ${{\bm{\bar{r}}}_{T,\alpha }} = \mathcal{R}\left[ \left( {{\bm{r}}_{T}}-{{\bm{x}}_{\alpha }} \right); -{{\theta }_{\alpha }} \right]$ represents the position vector of the $T$-th T1 transition relative to the $\alpha$-th topological defect; the observation period spans a duration $\tau$ preceeding the end of simulation (here $\tau = 200$).


\section{Thresholdless flows}

For $P_0 > P_0^{\ast} \approx 3.72$ (hexagonal cell pattern), we observe spontaneous tissue flows for very small cell activity, suggesting that the transition to flow is thresholdless, i.e. at exactly $\beta = 0$ in the limit of a very large system $N \rightarrow \infty$. 

\begin{figure}[h!]
\includegraphics[width=10cm]{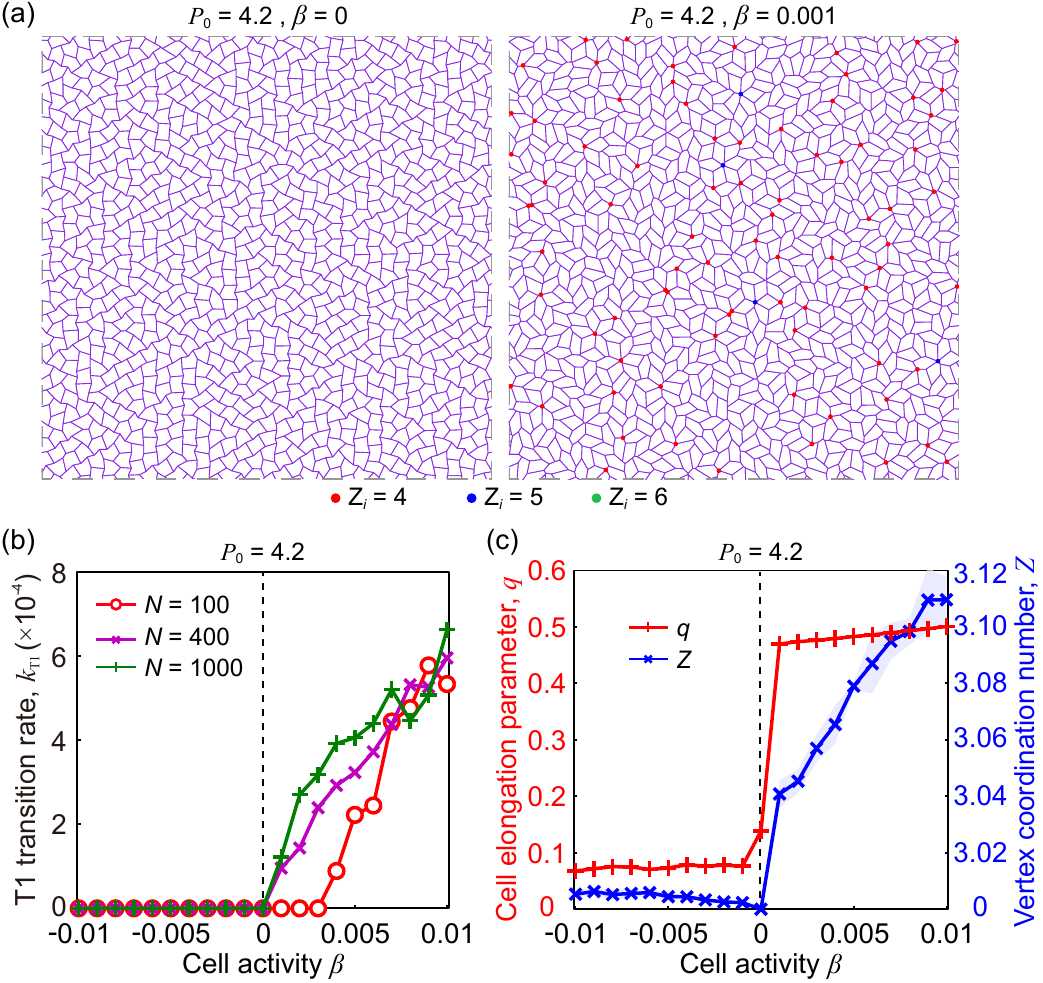}
\caption{\label{fig_ThresholdlessFlows} 
Thresholdless tissue flows for $P_0 > P_0^{\ast}$. 
(a) Tissue configuration at $\beta = 0$ (left) and $\beta = 0.001$ (right). 
(b) The T1 transition rate $k_{\rm T1}$ versus cell activity $\beta$ for different system sizes ($N = 100 , \ 400 , \ 1000$). 
(c) The cell elongation parameter $q$ and the average vertex coordination number $Z$ versus cell activity $\beta$. Here, system size $N = 1000$. 
Parameters: $K_P = 0.02$ and $P_0 = 4.2$. }
\end{figure}

For illustration, here we set $P_0 = 4.2$ and vary $\beta$. 
Figure \ref{fig_ThresholdlessFlows}(a) shows the morphology of a cell sheet at zero cell activity ($\beta = 0$) and small cell activity ($\beta = 0.001$), respectively. 
We observe strikingly different cell morphologies at $\beta = 0$ and $\beta = 0.001$: in the former, cells are concave polygons, with no rosettes; in the latter, cells are elongated $\beta = 0.001$, with many involved in rosettes.  In Fig. \ref{fig_ThresholdlessFlows}(b) we represent the T1 transition rate $k_{\rm T1}$ as a function of cell activity $\beta$ for different system sizes. 
Notably, when the system size is increased, the activity threshold for spontaneous T1 transitions shifts towards zero, which suggests thresholdless tissue flows. In addition, we observe a discontinuous (first-order like)  transition at $\beta = 0$ in both the cell elongation parameter $q$ and the average vertex coordination number $Z$ upon increasing cell activity $\beta$, as shown in Fig. \ref{fig_ThresholdlessFlows}(c). 
These results suggest that even a very small cell activity leads to a drastic change in the cell shape pattern (including the onset of multicellular rosettes).

\clearpage

\section{Supplemental Movies}

\textbf{Movie S1.} \textbf{Activity-induced melting} Tissue evolution under a gradually increasing cell activity, from $\beta = 0.17$ to $\beta = 0.50$. Here, we progressively increase the cell activity $\beta$ according to the step-by-step increase procedure: $\beta^{(n+1)} = \beta^{(n)} + \Delta \beta$, with $\Delta \beta = 0.003$; after each step increase in the cell activity, we relax the system for a time $t_{\rm relax} = 1,000$ (in $\gamma/K_A A_0$ unit). Parameters: $K_P = 0.02$ and $P_0 = 1$. 

\textbf{Movie S2.} \textbf{Hexagonal initial condition} Tissue evolution at different cell active stress levels ($\beta = 0.210, \ 0.325, \ 0.450$). Parameters: $K_P = 0.02$ and $P_0 = 1$. 

\textbf{Movie S3.} \textbf{Voronoi initial condition} Tissue evolution at different cell active stress ($\beta = 0.210, \ 0.325, \ 0.450$) levels for cells initially arranged according to a Voronoi pattern. Parameters: $K_P = 0.02$ and $P_0 = 1$.

\textbf{Movie S4.} \textbf{Four-cell system with free boundary} Evolution of a four-cell tissue under a gradually increasing cell activity, from $\beta = 0.1$ to $\beta = 0.5$. Here, we increase the cell activity $\beta$ step-by-step, $\beta^{(n+1)} = \beta^{(n)} + \Delta \beta$ with $\Delta \beta = 0.004$; after each increase in the cell activity, we relax the system for a time $t_{\rm relax} = 20,000$ (in $\gamma/K_A A_0$ unit). Parameters: $K_P = 0.02$ and $P_0 = 1$. 

\textbf{Movie S5.} \textbf{Single-cell system with free boundary} Evolution of a single-cell tissue under a gradually increasing cell activity, from $\beta = 0.1$ to $\beta = 0.5$. Here, we increase the cell activity $\beta$ step-by-step, $\beta^{(n+1)} = \beta^{(n)} + \Delta \beta$ with $\Delta \beta = 0.004$; after each increase in the cell activity, we relax the system for a time $t_{\rm relax} = 20,000$ (in $\gamma/K_A A_0$ unit). Parameters: $K_P = 0.02$ and $P_0 = 1$. 

\textbf{Movie S6.} \textbf{Four-cell system with periodic boundary} Periodic oscillations in a four-cell system under periodic boundary conditions. Parameters: $K_P = 0.02$, $P_0 = 1$, and $\beta = 0.315$. 

\textbf{Movie S7.} \textbf{Vertex-based inertia tensor} Tissue time evolution for different cell active stress levels using the vertex-based inertia tensor ($\beta = 0.250, \ 0.450, \ 0.800$). Parameters: $K_P = 0.02$ and $P_0 = 1$. 

\textbf{Movie S8.} \textbf{Area-based inertia tensor} Tissue time evolution for different cell active stress levels using the area-based inertia tensor ($\beta = 0.210, \ 0.250, \ 0.325$). Parameters: $K_P = 0.02$ and $P_0 = 1$.



%